\documentclass[journal]{IEEEtran}

\usepackage[pdftex]{graphicx}
\usepackage{epstopdf}
\usepackage{caption}
\usepackage{float}
\ifCLASSINFOpdf
   
\else
 
\fi
\usepackage[cmex10]{amsmath}
\begin{document}
%
\title{A Wideband Resistive Beam-Splitter Screen}
%
%
%
\author{NIVEDITA~MAHESH, RAVI~SUBRAHMANYAN, N.~UDAYA~SHANKAR, AND AGARAM~RAGHUNATHAN
\thanks{Manuscript received April, 2014.         } 
\thanks{The authors are with the Raman Research Institute, C V Raman Avenue, Sadashivanagar, Bangalore-560085, India (e-mail: nivedita@rri.res.in).}}

\markboth{AP1404-0496.R2}%
{Shell \MakeLowercase{\textit{et al.}}: Bare Demo of IEEEtran.cls for Journals}
%



\maketitle

\begin{abstract}

We present the design, construction and measurements of the electromagnetic performance of a wideband space beam splitter.  The beam splitter is a sheet in free space that is designed to divide incident radiation into reflected and transmitted components for interferometer measurement of spectral features in the mean cosmic radio background.  Analysis of a 2-element interferometer configuration with a vertical beam splitter between a pair of antennas leads to the requirement that the beam splitter be a resistive sheet with sheet resistance $\eta_o/2$, where $\eta_o$ is the impedance of free space.  The transmission and reflection properties of such a sheet are computed for normal and oblique incidences and for orthogonal polarizations of the incident electric field.  We have constructed such an electromagnetic beam splitter as a square soldered grid of resistors of value 180~Ohms (approximately $\eta_o/2$) and a grid size of 0.1~m.  We measured the reflection and transmission coefficients over a wide frequency range between 50 and 250~MHz in which the wavelength well exceeds the mesh size.  Measurements of the coefficients for voltage transmission and reflection agree to within 5\% with physical optics modeling of the wave propagation, which takes into account edge diffraction.

\end{abstract}



%
\IEEEpeerreviewmaketitle

\section{Introduction}

%
%
%
%


\IEEEPARstart{E}{vents} in the cosmological evolution of baryons in the Universe are predicted to have left imprints in the spectrum of the cosmic radio background.  The primordial gas was predominantly Hydrogen.  At cosmic times between 0.13 to 1.35 Myr the expansion of the Universe resulted in a transformation of the state of the Hydrogen from highly ionized to almost completely neutral.  In this Recombination Epoch the trickle of bound electrons down atomic states adds recombination lines to the radiation content.  This forest of recombination lines is expected to be detectable today as a ripple in the spectrum of the cosmic radio background [1,2].  Subsequently, following the formation of the first collapsed objects---which may be massive Population III stars or ultra-dwarf galaxies---the neutral recombined hydrogen gas is irradiated by first light.  This energetic UV and X-ray radiation transforms the level populations in the Hydrogen 21-cm spin flip transition, alters the strength of the coupling between these levels and the kinetic temperature of the gas.  It also heats the gas thus altering the kinetic temperature and finally results in a reionization of the intergalactic hydrogen.  The interaction between these spin-flip levels with the cosmic microwave background changes over cosmic time resulting in spectral distortions at frequencies below 200~MHz where the redshifted 21-cm transition appears [3].

These spectral features caused by events during the epochs of recombination and reionization as well as relatively wider spectral distortions in the cosmic microwave background that arise from energy injection in the early universe---followed by partial or saturated Comptonization---are all orders of magnitude smaller in brightness compared to the brightness of the radio sky. Therefore, their detection requires development of purpose-built sensitive spectral radiometers.  Previous work in this direction includes proposals for antenna elements [4] and receiver system designs [5] as well as space missions [6]. These have been for single element total power spectral radiometers.  Some systems have been deployed in remote sites; EDGES [7], SCI-HI [8], BIGHORNS [9] and SARAS [10] have reported results of observations. Some of these have yielded outcomes of scientific value; however, they have all been limited by systematic errors and methods that reduce these are critical to progress in this research.

A limiting factor in absolute measurements of the spectrum of the sky with a spectral radiometer is unwanted additive contributions generated within the receiver electronics.  Radiometers to date rely on switching schemes to cancel these.  Most often the input to the receiver is switched between the sensor of the sky electromagnetic radiation and a load that provides a reference spectrum.  Another common technique is the use of correlation receivers in which the signal from the sensor is split into two parts, amplified in separate receiver chains and a cross correlation spectrum computed between the two arms.  This nominally eliminates the receiver noise contributions because their noise voltages are statistically independent. 

Absolute measurements of the cosmic radio background using a single antenna element as the sensor of the sky radiation will always include emissions from the resistive losses in the antenna and associated balun.  Additionally, although switching schemes and correlation receivers do cancel internal sources of receiver noise to a large extent, accuracy in spectral measurements is limited by small-magnitude changes in system bandpass, impedance matching in switching schemes and multi-path propagation of internal noise.   These manifest in the measurement as uncalibrated spectral structure that often takes the form of spectral ripples with a range of periods.

Radio interferometers have the advantage of being unresponsive to additive noise contributions arising from ohmic loss in antennas as well as noise in amplifiers and receiver chains.  This is because these components, which are generated in different arms of an interferometer, are uncorrelated. However, interferometers are also almost completely insensitive to the uniform sky brightness and hence unsuitable for absolute measurement of the cosmic radio background and its spectrum.  A way to retain the advantages of interferometers and make absolute measurements of spectral structure in the uniform sky brightness is to construct a space beam splitter that divides and directs the incident sky power into two antennas that form an interferometer pair.  The interferometer then responds to, and provides a measurement of, the common mode power.  We present below this concept, which we call a `zero-spacing interferometer'.  The sections that follow develop the design of the electromagnetic space beam splitter for this configuration.  The system design of the zero-spacing interferometer will be presented in a subsequent manuscript.

\section{The concept of a zero-spacing interferometer}

\begin{figure}[ht] 
\centering
\includegraphics[width = 3.4in]{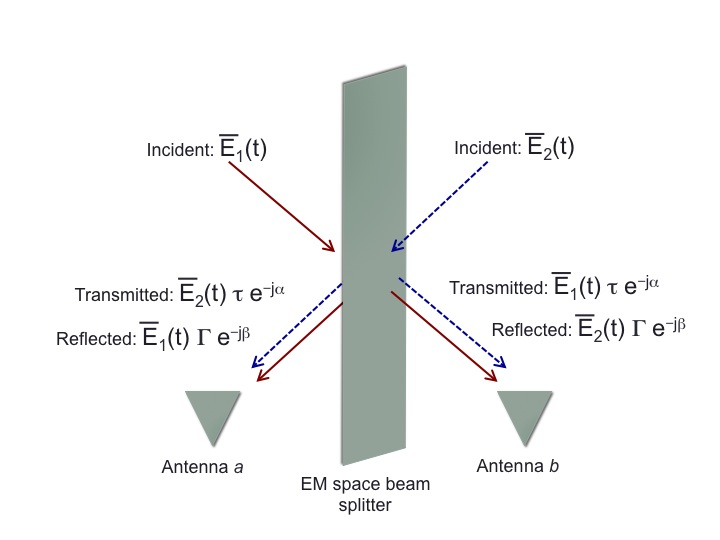}
\caption{Configuration of the zero-spacing interferometer showing the placement of the space beam splitter vertically between the pair of antennas.  The presence of the beam splitter causes the interferometer to respond to the mean sky brightness.}
\label{fig:screen}
\end{figure}

The configuration of a zero-spacing interferometer that responds to the mean cosmic radio background is shown in Fig.~\ref{fig:screen}.  A vertical sheet that is an electromagnetic beam splitter is placed at the geometric center of a pair of antennas, which are placed at a common height above ground.  The antennas are frequency independent and their radiation patterns are directed on the two sides of a common portion of the vertical sheet.   The ground below the antennas and vertical sheet is covered with absorber tiles to make the antenna pattern on the sheet independent of frequency.  Radiation from throughout the sky above is incident on the vertical sheet from either side.  Since the sheet is a beam splitter, all incident radiation is partially transmitted and partly reflected. Each antenna, therefore, receives sky EM radiation transmitted through the beam splitter from the far side as well as reflected off the beam splitter from the near side.  The fields sensed by the antennas are amplified and cross-correlated and the spectral distribution of the cross power spectrum is a measurement of the spectral distribution of the absolute sky brightness.  The antenna pair forms an interferometer that is potentially sensitive to uniform sky brightness; therefore, we refer to this configuration as a `zero-spacing interferometer'.  It is of interest to first compute the transmission and reflection properties of the beam splitter that would maximize the sensitivity to this component.

As shown in Fig.~\ref{fig:screen}, $\overline{E_1(t)}$ and $\overline{E_2(t)}$ represent the EM waves from the sky that are incident on the two sides of the beam splitter and $\Gamma e^{-j\beta}$ and $\tau e^{-j\alpha}$ represent the complex reflection and transmission coefficients.  The signals sensed by each of the two antennas, $\overline{E_a(t)}$ and $\overline{E_b(t)}$, are vector sums of reflected and transmitted signals:
\begin{equation}
\overline{E_a(t)} = \overline{E_{1}(t)}\Gamma e^{-j\beta} + \overline{E_{2}(t)}\tau e^{-j\alpha}, 
\end{equation}
and
\begin{equation}
\overline{E_b(t)} = \overline{E_{2}(t)}\Gamma e^{-j\beta} + \overline{E_{1}(t)}\tau e^{-j\alpha}.
\end{equation}
Cross-correlation of these antenna signals gives the measurement:
\begin{eqnarray}
\langle \overline{E_a(t)}\  \overline{E_b(t)}^{\ast} \rangle  &=&  \langle |E_{1}(t)|^2 \Gamma \tau e^{-j(\beta- \alpha)} \rangle \nonumber \\
&& + \langle |E_{2}(t)|^2 \Gamma \tau e^{-j(\alpha-\beta)} \rangle \nonumber\\
&& + \langle \overline{E_1(t)}\ \overline{E_2(t)}^{\ast} | \Gamma |^2 \rangle \nonumber \\
&& + \langle \overline{E_1(t)}^{\ast}\ \overline{E_2(t)} | \tau |^2 \rangle.
\end{eqnarray}
Since the two incident beams $\overline{E_1(t)}$ and $\overline{E_2(t)}$ are from different sky directions they are uncorrelated and hence the last two terms in the above equation average to zero.  The product becomes
\begin{eqnarray}
\langle \overline{E_a(t)}\  \overline{E_b(t)}^{\ast} \rangle & = &
\Gamma \tau (| \overline{E_{2}(t)} |^2 + | \overline{E_{1}(t)} |^2) {\rm cos}(\beta - \alpha) \nonumber \\
&& - j \Gamma \tau (| \overline{E_{2}(t)} |^2 - | \overline{E_{1}(t)} |^2)  \nonumber \\
&& {\rm sin}(\beta - \alpha)
\label{equal_split}.
\end{eqnarray}
Equation~\ref{equal_split} is the measurement equation for the zero-spacing interferometer; the real part is proportional to the sum of the powers incident from the two sides of the beam splitter where as the imaginary part is proportional to the difference.  For uniform sky, the imaginary component will vanish.

The pair of antennas for the zero-spacing interferometer will have to be directive and with substantial fraction of their beam solid angle on the screen.  Nevertheless, because the antenna patterns will also have sidelobes directly on the sky, there will be direct rays from the sky to the antennas and that will yield unwanted responses.   In the planned experimental setup, the screen is oriented North-South and so as the sky drifts East to West every part of the sky results in a response from both East and West sides of the screen as the piece of sky transits the meridian.  Therefore, when the response is averaged over 24~h in local sidereal time, the spectral structure in the interferometer response that is due to direct rays will average over time to give a constant signal across the observed spectrum.  What will survive is the spectral structure due to spectral signals in the cosmic radio background, which is what the zero-spacing interferometer is designed to measure. In the average response, this will be the real part of Equation~\ref{equal_split}.

\section{Response of a zero-spacing interferometer with a lossless space beam splitter}

For a beam splitter that has no charge or time varying magnetic field and hence no curl or divergence sources of the electric field, the electric field will be continuous across the surface.  This implies that
\begin{equation} 
1 + \Gamma \cdot e^{-j\beta} = \tau\cdot e^{-j\alpha}.
\end{equation} 
The real and imaginary parts of this equation lead separately to the relationships: $(1 + \Gamma {\rm cos}\beta) = \tau {\rm cos} \alpha$ and $\Gamma {\rm sin}\beta = \tau {\rm sin} \alpha$.  For a lossless electromagnetic beam splitter, conservation of power requires that $\Gamma^{2} + \tau^{2} = 1$.  These lead to the condition that $(\beta - \alpha) = \pi/2$, which implies that for such a lossless beam splitter the measurement $\langle \overline{E_a(t)}\  \overline{E_b(t)}^{\ast} \rangle$ has zero response to a uniform sky background. 

\begin{figure}[!h] 
\centering
\captionsetup{justification=centering}
\includegraphics[width = 3.3in]{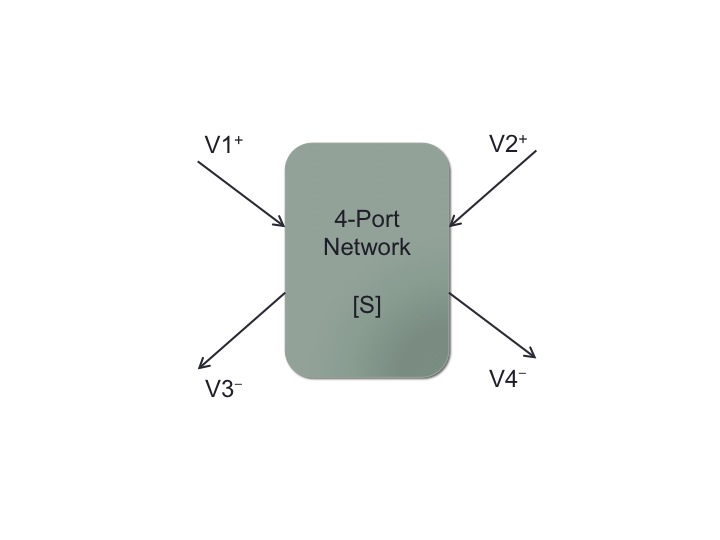}
\caption{A four-port network model for any composite space beam splitter.}
\label{fig:four_port}
\end{figure}

Thus, beam splitters that are capacitive or inductive and hence lossless would have net phase difference of $\pi/2$ between reflected and transmitted components and for such beam splitters the net response of the zero-spacing interferometer would be zero for uniform sky (see Equation~\ref{equal_split}).  We next consider the case of beam splitters that are composed of multiple parallel sheets, which may be individually capacitive or inductive, but whose spacings are adjusted to tune the net phase.  Such a compound beam splitter may be described by it its net properties as a four-port network of the form shown in Fig.~\ref{fig:four_port}, with net transmission and reflection described using scattering parameters. 

As shown in Fig.~\ref{fig:four_port}, the incoming waves $[V^+]$ and outgoing waves $[V^-]$ are related by the scattering matrix [S] as:
\begin{equation} 
[V^-] = [S]\cdot[V^+].
\end{equation}

For a four-port lossless reciprocal network, the relevant terms in the scattering matrix are
\begin {equation}
\left[ \begin{array} {c}
V1^-\\V2^-\\V3^-\\V4^- \end{array}\right] = \left[ \begin{array} {cccc} 0 & 0 & S_{13} &S_{14}\\ 0 & 0 & S_{23} &S_{24}\\S_{31} & S_{32} & 0 &0\\S_{41} & S_{42} & 0 &0 \end{array}\right] \cdot \left[ \begin{array} {c} V1^+\\V2^+\\V3^+\\V4^+ 
\end{array}\right].
\end{equation}
The two outputs, $V3^-$ and $V4^-$ can be written as:
\begin{equation} V3^- = S_{31}~V1^+ + S_{32}~V2^+   \end{equation}
and
\begin{equation} V4^- = S_{41}~V1^+ + S_{42}~V2^+. \end{equation}

The scattering matrix [S] for any lossless reciprocal network is symmetric and unitary[11], which implies that
\begin{equation}
S_{13} = S_{31},~  S_{23} = S_{32},~ S_{14} = S_{41}, S_{24} = S_{42},
 \end{equation} 
 \begin{equation}
\sum_{k=1}^4 S_{ki}S_{ki}^{\ast}  =  1 ~~~~\forall~~ i
\end{equation}   
and
 \begin{equation}
\sum_{k=1}^4 S_{ki}S_{kj}^{\ast}  =  0 ~~~~\forall~~ i\neq j.
\end{equation}   
From the above conditions we infer that
\begin{equation} 
~S_{31} S_{41}^{\ast}  =  - S_{32} S_{42}^{\ast}.
\label{eqn_11} 
\end{equation} 
This result may be written in terms of the net transmission and reflection coefficients as
\begin{equation} 
\Gamma e^{-j\beta} \tau e^{j\alpha} = - \tau e^{-j\alpha} \Gamma e^{j\beta}.
\end{equation}
This leads to the result that $ (\alpha - \beta) = \pi/2$, or that even for a compound beam splitter composed of lossless capacitive or inductive sheets the measurement $\langle \overline{E_a(t)}\  \overline{E_b(t)}^{\ast} \rangle$ yields a null value for uniform sky.

A lossless space beam splitter, even if composite, will manifest a phase difference of $\pi/2$ between the transmitted and reflected components as is the case for a partially silvered mirror.  The lack of response to uniform sky is a consequence of this property.  This leads to the conclusion that if the zero-spacing interferometer should respond to uniform sky, the beam splitter has to be lossy and not have a $\pi/2$ phase shift between transmission and reflection coefficients.

\section{EM propagation through a beam splitter with finite sheet conductivity}

The conductance $G$ between a pair of opposite faces of a uniform rectangular slab of conducting material is given by 
\begin{equation}
G = \sigma  \frac{A}{l},
\end{equation}
where $A$ is the surface area of each of the two faces, $l$ is the spacing between the faces and $\sigma$ is the bulk conductivity of the material of the slab.  Sheet conductance ($G_s$) of a conducting sheet is defined as the conductance between a pair of opposite edges of a square piece of this sheet.  If the sheet has thickness $t$, $A = lt$, and
\begin{equation}
G_s = \sigma t.
\end{equation}
The unit for sheet conductance is Siemen-square.

The space beam splitter is taken to be a sheet of conductance $S=\sigma \cdot \delta x$ (Siemen-square), where $\sigma$ is the material conductivity and $\delta x$ is the sheet thickness.  When an electromagnetic wave traveling in one medium is incident on a second medium with a different intrinsic impedance, in general the wave will be partially reflected and partially transmitted. Since the second medium considered here has a definite value of conductance, a part of the incident energy will also be absorbed at the sheet.

\subsection{Transmission and Reflection coefficients}

We adopt right handed Cartesian coordinates with the sheet placed in the $yz$-plane. The incoming radiation is obliquely incident on the sheet.  $E_i$ denotes the electric field intensity of the incident wave, $E_r$ denotes the electric field intensity of the reflected wave and $E_t$ denotes the electric field intensity of the transmitted wave on the far side of the sheet. Similar subscripts are used to denote the magnetic field intensities of the corresponding waves. The electric and magnetic field components are written with a negative sign if their corresponding vector directions are opposite to that of the Cartesian axes. The Poynting vector for the incident wave is assumed to be in the $xz$-plane and at an angle $\theta$ (the angle of incidence) to the sheet normal. Snell's law applies here and the angle of reflection is equal to that of incidence. The angle of transmission is the same as angle of incidence because the wave travels into the same medium (air, whose intrinsic impedance $\eta_o$ is 377~$\Omega$) following propagation through the sheet. For oblique incidence it is necessary to consider a pair of polarizations for the incident electric field: (1) an E-plane incidence, where the incident E-field is in the plane of incidence and (2) an H-plane incidence, where the incident E-field is perpendicular to the plane of incidence.  
	
The boundary conditions for a conductive sheet are:
\begin{itemize}
\item the discontinuity in the normal component of $E$ is proportional to the surface charge density,
\item the normal component of $B$ is continuous,
\item the tangential component of $E$ is continuous, and
\item the discontinuity in the tangential component of $H$ equals the surface current density.
\end{itemize}	

\begin{figure}[ht] 
\centering
\includegraphics[width = 3.4in]{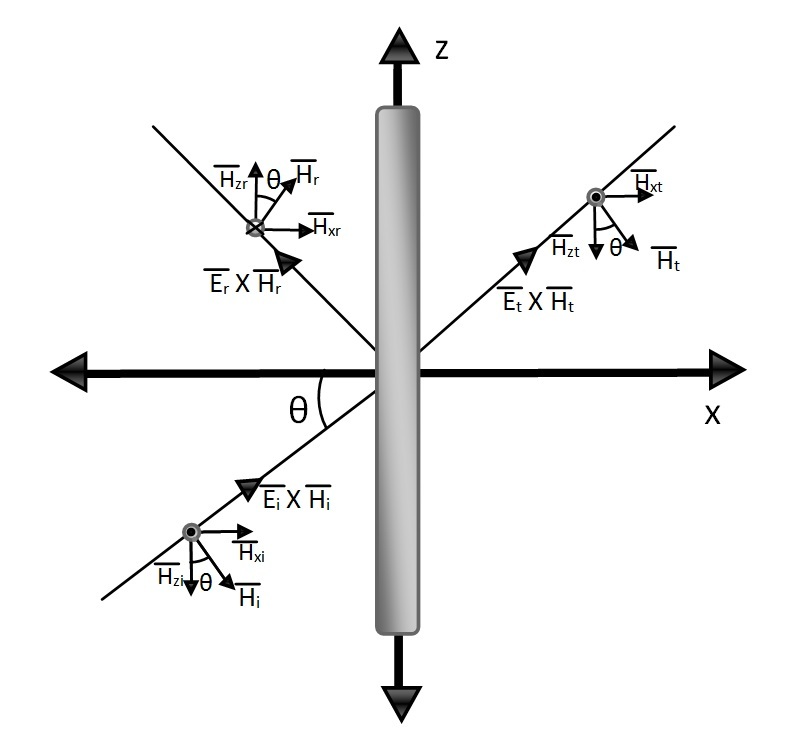}
\caption{The orientations of E and H fields of the incident, reflected and transmitted EM waves for the case where the E-field of the incident radiation is polarized perpendicular to the plane of incidence.  The angle of incidence is $\theta$ and the plane of incidence is the $xz$-plane.}
\label{fig:EM_||}
\end{figure}

We first consider the case of H-plane incidence.  The incident, reflected and transmitted waves and the orientations of their respective components $E$ and $H$ for this case are shown in Fig.~\ref{fig:EM_||}. $\overline{H_i}$ is at angle $\theta$ (the angle of incidence) to the $z$-axis and this field may be resolved into two perpendicular components $H_{xi}$ and $-H_{zi}$. $\overline{E_i}$, $\overline{E_r}$ and $\overline{E_t}$ are all along negative $y$-axis.  

There are no static charges on the sheet and currents driven by the electric field are along y-axis.  We define $\overline{J}$ to represent the vector current along positive $y$ axis.  

We denote the electric field at the sheet by $E_c$.  We may substitute $S \times E_c$ for $-J_y \times \delta x$ and write the boundary condition for tangential H field in terms of $E$ fields to get:
\begin{equation} \frac{1}{\eta_o} (E_i \rm{cos}\theta - E_r\rm {cos}\theta - E_t \rm{cos} \theta) = S E_c.\end{equation}
Using the condition that this electric field be continuous, we may rewrite this expression in the form
\begin{equation} E_r = - \frac{\eta_o S E_c}{2\rm{cos}\theta}.\label{E_r} \end{equation}
The net electric field $E_c$ along the sheet is given by the sum of $E_i$ and $E_r$:
\begin{equation} E_c = E_i - \frac{\eta_o S E_c}{2\rm{cos}\theta}.\end{equation}
We may now substitute the above expression for $E_c$ in Equation~\ref{E_r} to derive the reflection coefficient, which is the ratio of reflected to incident electric fields:
\begin{equation} \Gamma =  \frac{\eta_o S}{2\rm{cos}\theta + \eta_o S} \angle 180^{\circ}. \label{gamma_H}\end{equation}
The transmission coefficient, which is the ratio of transmitted to incident electric fields, may be evaluated from the relation $1 + \Gamma = \tau$ that follows from applying boundary conditions to integral forms of Maxwell's equations.  We derive that
\begin{equation} \tau = \frac{\rm{cos}\theta}{\rm{cos}\theta + \eta_o S/2}\angle 0^{\circ}.\label{tau_H} \end{equation}

If the incidence angle $\theta$ is set to be $0^o$ (this is the case of normal incidence), $\Gamma$ and $\tau$ reduce to the forms
\begin{equation} \Gamma =  \frac{\frac{\eta_o S}{2}}{1 + \frac{\eta_o S}{2}} \angle 180^{\circ}\end{equation}
and
\begin{equation} \tau = \frac{1}{1 + \frac{\eta_o S}{2}}\angle 0^{\circ}. \end{equation}

\begin{figure}[H] 
\centering
\includegraphics[width = 3.4in]{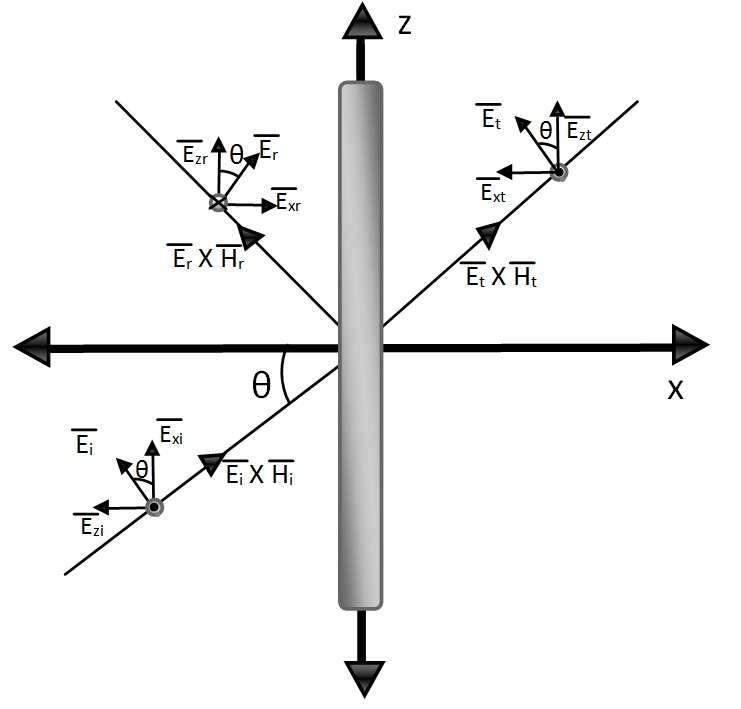}
\caption{The orientations of E and H components of the incident, reflected and transmitted EM waves when the incident wave is polarized to have its electric field in the plane of incidence.}
\label{fig:EM_perpen}
\end{figure}

We next consider the case of E-plane incidence. For an incoming wave with electric field polarized to be in the plane of incidence, the electric field will make an angle $\theta$ (equal to the angle of incidence) to the $z$-axis. The $E$ components of the incident, reflected and transmitted waves may be resolved into $E_x$ and $E_z$ components as shown in Fig.~\ref{fig:EM_perpen}.

We once again use the integral forms of Maxwell's equations, along with boundary conditions, to derive the change in fields across the sheet.  We assume that there are no net charges in the resistive sheet.   The net current density is assumed to reside within the sheet and it would be in the plane of incidence, since that is the direction of the electric field.  

We have taken the current density vector $\overline{J}$ in this case study to be along positive $z$-axis.

The condition that the tangential H field be continuous may be written in terms of the electric field components as follows:
\begin{equation} \frac{1}{\eta_o} (E_{i} - E_{r} - E_{t}) = S \times E_c,  \end{equation}
where $\overline{E_c}$,  the vector electric field in the sheet in this case, is also along positive $z$-axis.
The angles of incidence, reflection and transmission are equal and a joint solution of the above equation together with the electric field continuity equation yields:
\begin{equation} E_{r} = -\frac{\eta_o S E_c}{2}.\label{E_xr}\end{equation}

The net electric field within the sheet is due to the $z$- components of the electric fields of the incident and reflected waves:
\begin{equation} E_c = E_{zi} -\frac{\eta_o S E_c \rm{cos}\theta}{2}.\end{equation}
Substituting this expression for $E_c$ in Equation~\ref{E_xr} leads to an expression for the ratio of reflected to incident electric fields, which is the reflection coefficient:
\begin{equation} \Gamma = \frac{\frac{\eta_o S}{2} \rm{cos}\theta}{1+\frac{\eta_o S}{2} \rm{cos}\theta} \angle 180^{\circ}.\label{gamma_E}\end{equation}
The transmission coefficient is then calculated to be:
\begin{equation} \tau = \frac{1}{1+\frac{\eta_o S}{2} \rm{cos}\theta} \angle 0^{\circ} .\label{tau_E}\end{equation}

For the case of normal incidence, we substitute $\theta = 0^o$ in the above expressions for $\Gamma$ and $\tau$ to get:
\begin{equation}\Gamma = \frac{\frac{\eta_o S}{2}}{1+\frac{\eta_o S}{2}} \angle 180^{\circ} \label{gamma_theta} \end{equation}
and
\begin{equation} \tau = \frac{1}{1+\frac{\eta_o S}{2}} \angle 0^{\circ} \label{tau_theta}. \end{equation}
For normal incidence, the expressions for reflection and transmission coefficients are the same for E and H plane incidence.

\subsection{The desired impedance of a space beam splitter in a zero-spacing interferometer}

We have chosen the sheet impedance of the electromagnetic beam splitter to be the value that would maximize the response of the zero-spacing interferometer to uniform sky.  In the measurement equation (Equation~\ref{equal_split}), the real component is proportional to the sum of the sky radiation that is incident on the two sides of the beam splitter.  Therefore, the sheet impedance is chosen to maximize the real part of the measurement equation, ignoring the imaginary component. 

For normal incidence, the reflection and transmission coefficients (Equations~\ref{gamma_theta} and \ref{tau_theta}) have a phase difference of $180^o$.  The response to uniform sky is a maximum when cos($\beta - \alpha$) is unity, which requires that the conductance of the space beam splitter should be a real value, {\it i.e.}, the sheet is purely resistive.  

The value of $S$ should also be such that it maximizes the product:
\begin{equation} \Gamma \times \tau = \frac{S\frac{\eta_o}{2}}{(1+S\frac{\eta_o}{2})^2},  \end{equation}
which is a maximum when $S = 2/\eta_o$. At this point of inflection the magnitudes of the reflection and transmission coefficients for normal incidence become equal and equal to $1/2$.  Therefore, maximizing the product and thereby maximizing the real term in Equation~\ref{equal_split} leads to the result that the sheet conductance $S$ is required to be real and  equal to $2/\eta_o$~Siemen-square.  

For such a resistive sheet, the power reflected, transmitted and absorbed (per unit area) may be expressed in terms of the incident power (per unit area) by the following expressions:
\begin{equation} {\rm Incident~power} = | \overline{E_i} \times \overline{H_i} | =  \frac{E_i^2}{\eta_o}, \end{equation}
\begin{equation} {\rm Reflected~power} = \frac{E_r^2}{\eta_o}  =\frac{1}{\eta_o} \left(S\frac{\eta_o}{2}\right)^2 \left(\frac{E_i}{1+S\eta_o/2}\right)^2 \nonumber \end{equation}  
\begin{equation} = \frac{1}{4} \times  \frac{E_i^2}{\eta_o}, \end{equation}
and
\begin{equation} {\rm Transmitted~power} = \frac{E_t^2}{\eta_o} = \frac{1}{\eta_o} \left(\frac{1}{1+S\frac{\eta_o}{2}}\right)^2  = \frac{1}{4} \times  \frac{E_i^2}{\eta_o}. \end{equation}
The power absorbed by the resistive sheet (per unit area) may be calculated as the ratio of ohmic dissipation and current density; this is given by:
 \begin{equation} {\rm Power~absorbed} = (J \times \delta x)^2/S = E_c^2 \times S = E_i^2/(2 \eta_o). \end{equation} 
Thus, half the incident power is absorbed by the resistive sheet, a quarter is reflected and the remaining quarter is transmitted.

\subsection{Dependence of the interferometer response on the angle of incidence }

\begin{figure}[H] 
\centering
\includegraphics[width = 3.4in]{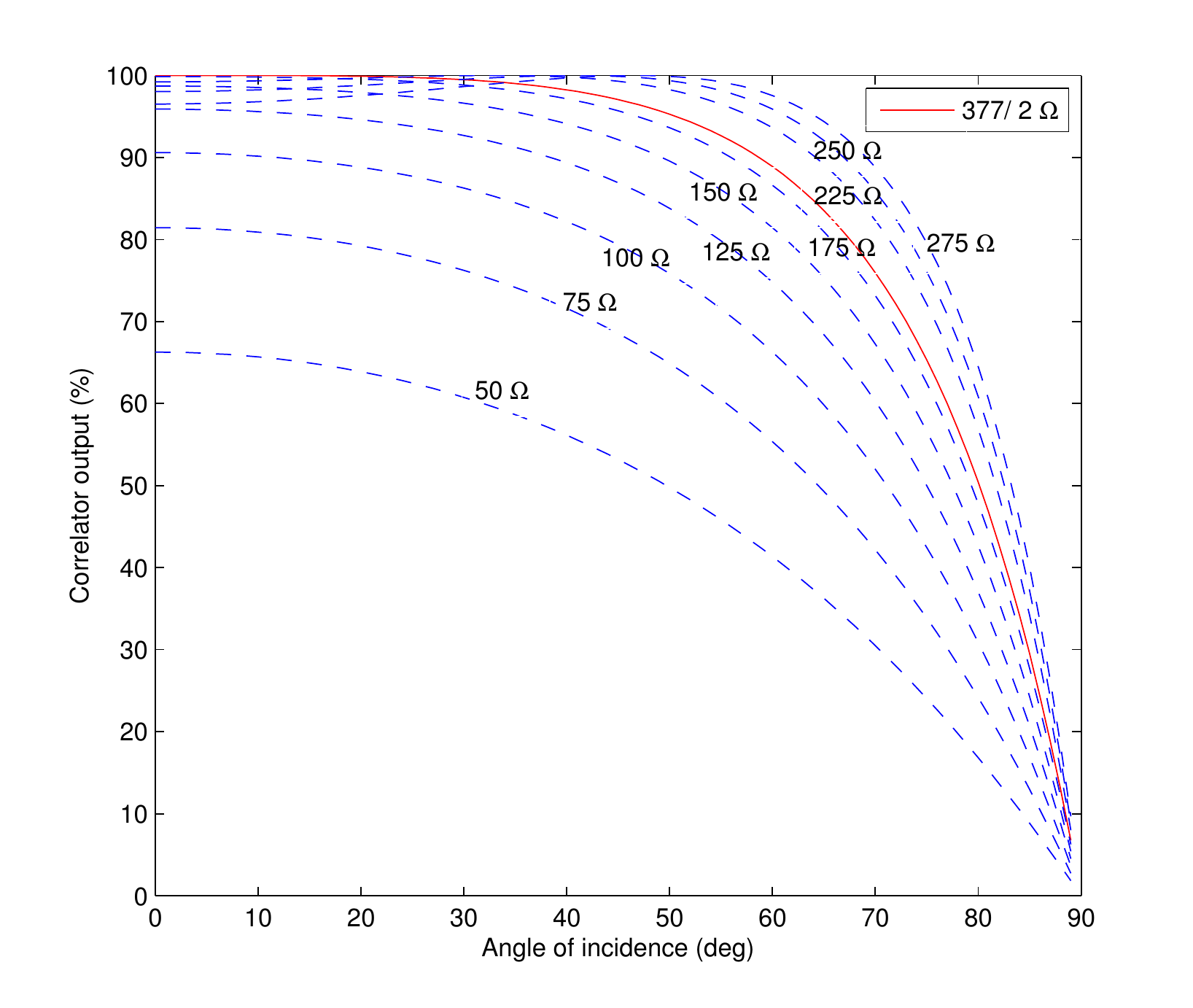}
\caption{ Response of the zero-spacing interferometer to uniform sky, versus angle of incidence.  The normalized response is computed as the ratio of $\Gamma \tau$ to the maximum value of 0.25 that this product may take.  Different traces are for different values of sheet conductance $S$.  Shown as a continuous line is the curve corresponding to $S = 2/\eta_o$, or a sheet resistance of 188.5~$\Omega$~square$^{-1}$.  The three lines above the continuous line are, progressively, for sheet resistance $(1/S)$ of values 225 to 275~$\Omega$~square$^{-1}$ in increments of 25.   The six lines below the continuous line are for decreasing values of sheet resistance from 175 to 50~$\Omega$~square$^{-1}$ again in intervals of 25.}
\label{fig:angle_sensitivity}
\end{figure}

The transmission and reflection coefficients have a dependence on the angle of incidence $\theta$.  For a space beam splitter that is a resistive sheet, these coefficients are real quantities and their values approach $1/2$ as $\theta$ tends to $0^{\circ}$, which is the case of normal incidence.  Therefore, when such a resistive sheet is deployed as the space beam splitter in a zero-spacing interferometer, the magnitude of response to the uniform sky brightness would depend on incidence angle via the $\Gamma \tau$ product term in Equation~\ref{equal_split}.  In Fig.~\ref{fig:angle_sensitivity} we plot the magnitude of this response, normalized to the maximum value of 0.25 that this product might take for normal incidence, for different values of sheet conductivity.  

For normal incidence, the response is a maximum for $S = 2/\eta_o$, which corresponds to a sheet resistance of 188.5~$\Omega$~square$^{-1}$.  For larger values of sheet resistance the response is a maximum at progressively larger angles of incidence.  A sheet with conductance $S$ would present an effective conductance of $S/{\rm cos}\theta$ to radiation with angle of incidence $\theta$.  The transmission and reflection for radiation with angle of incidence $\theta$ corresponds to this effective conductance that the sheet presents for radiation incident at angle $\theta$.  As seen in Fig.~\ref{fig:angle_sensitivity}, as the sheet resistance increases above $\eta_o / 2~\Omega$~square$^{-1}$, the response is a maximum at that angle for which $S/{\rm cos}\theta$ equals $2 / \eta_o $.  A value somewhat exceeding $\eta_o / 2~\Omega$~square$^{-1}$ would provide a good response over a wide range in incidence angle, the appropriate value would depend on the beam patterns of the antennas and the geometry of the zero-spacing interferometer.

\subsection{Considerations arising from skin depth of the sheet material}

At frequency $f$, the skin depth $\delta_s$ is given by:
\begin{equation}
\delta_s = 1/\sqrt{\pi \sigma f \mu_o \mu_r},
\end{equation}
where $f$ is the frequency of the incident EM wave and $\mu_o$ and $\mu_r$ are, respectively, the permeability of free space and the relative magnetic permeability of the sheet material.
At frequencies where the skin depth $\delta_s$ becomes comparable to or less than the sheet thickness $\delta x$, the sheet conductance falls below the d.c. conductance $\sigma \times \delta x$ and is given by
\begin{equation}
S = \sigma \times \delta x \{ 1 - e^{-\delta x/\delta_s}\}.
\end{equation}
In general, it is this a.c. conductance that is required to be $\eta_o/2$, which implies that
\begin{equation}
\sigma \times \delta x \{ 1 - e^{-\delta x/\delta_s}\} = \eta_o/2
\end{equation}
for operation as a beam splitter.  

The a.c. conductance of the sheet tends to $\sigma \times \delta_s$ for sheets of large thickness and to $\sigma \times \delta x$ for the case when the sheet thickness is much smaller than the skin depth.  Since we seek a frequency independent performance for the beam splitter sheet, the above analysis leads us to conclude that the screen should be thin, which implies that 
the product of bulk conductivity and the square of the sheet thickness must satisfy the relation:
\begin{equation}
\sigma \times (\delta x)^2 \ll 1/(\pi f  \mu_o \mu_r)
\end{equation}
over the entire frequency range of operation.  If we assume that the relative magnetic permeability of the sheet material is unity, this leads to the result that the sheet thickness needs to be:
\begin{equation}
\delta x \ll (6.913/f_{\rm 100MHz})~{\rm cm},
\end{equation}
where $f_{\rm 100MHz}$ is the frequency in units of 100~MHz: $f_{\rm 100{\rm MHz}} = f/(100~{\rm MHz})$.

\subsection{Mutual coherence in the emissivity on opposite sides of a resistive sheet: an additive response in the zero-spacing interferometer}

There will be a partial mutual coherence between the thermal emission that emerges on the two sides of a resistive sheet.  The emissivity will depend on the physical temperature of the resistive sheet and its opacity.  The fractional coherence between the emission on the opposite sides will depend on the opacity: an opaque screen will have maximum emissivity and zero mutual coherence and as the opacity reduces the emissivity reduces and the fractional mutual coherence increases.  If the sheet conductance $S$ is real and  equal to $2/\eta_o$~Siemen-square, then half the incident sky radiation would be absorbed in the sheet, reducing the signal that arrives at the sensors.  Additionally, sheet emission emerges normally on the two sides with intensities that correspond to a brightness temperature that is equal to half the physical temperature of the sheet.  Both these result in an increase in the measurement noise or error.  A resistive beam splitter results in a reduction of the signal to noise ratio; however, this is unavoidable because as we have shown above, a zero-spacing interferometer with a lossless beam splitter would have zero response to uniform sky.  At long wavelengths the sky brightness temperature of the cosmic radiation is greater than the physical temperature of the beam splitter and hence the measurement noise contributed by the sheet emission would not be dominant. At long wavelengths the main drawback of using a resistive beam splitter is the factor of two loss in signal and hence about a factor of two loss in sensitivity.  These reductions in sensitivity need to be traded off against the advantage in terms of reduced systematics that accrues from using the proposed configuration.  Precision measurements of the radio background suffer from systematics, which can be a show stopper, and progress in making sensitive measurements comes from new methods that cancel and avoid systematics rather than increase the raw sensitivity.   




\section{A resistive wire mesh or square grid of resistors functioning as a resistive sheet}

A resistive sheet may be approximated by a square wire grid made of resistive wire in which the grid size is considerably smaller than the wavelength of the EM waves.  Such a realization has been used previously for the construction of an absorber screen [12]. Frequency independent operation as a beam splitter requires that the wire radius be much smaller than the skin depth over the frequency range of operation, so that the a.c. conductance of the wires is the same as the d.c. conductance.  The sheet resistance of a wire grid that has square grids is equal to the resistance value of a single wire segment [12]; therefore, every wire segment of the grid is to be made equal to $(\eta_o/2)~\Omega$ for operation as a beam splitter.

Alternately, the resistive sheet may be constructed as a square grid of resistors.  The sheet resistance of a square grid of resistors is equal to the value of the individual resistors, which implies that our beam splitter resistive sheet may be constructed as a grid of resistors in which every resistor has a value of $(\eta_o/2)~\Omega$.  It is assumed here that the resistors have a.c. resistance equal to this value over the entire frequency range of operation.

\subsection{Performance dependence on the size of the resistor grid}

\begin{figure}
\centering
\includegraphics[width = 3.4in]{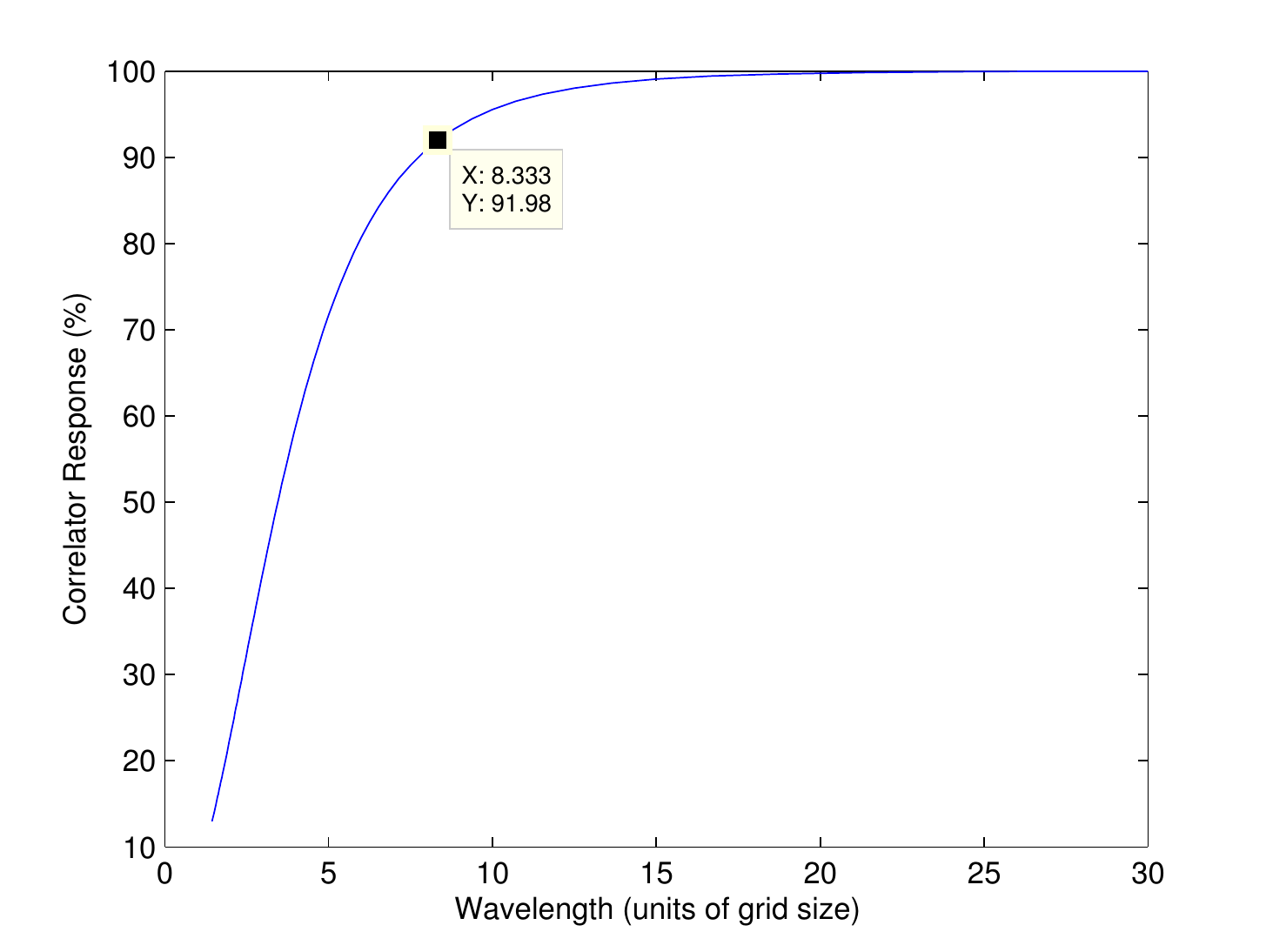}
\caption{Frequency dependence of the response of the zero-spacing interferometer to uniform sky brightness when the resistive sheet is implemented as a square grid of resistive wire or resistors.  The response is shown as percentage of the maximum response versus the operating wavelength, which is in units of the grid size. The response is computed as $\Gamma \tau$ product normalized to the maximum value of 0.25.}
\label{fig:graph_gridloss}	
\end{figure}

The EM space beam splitter constructed from a grid of resistive wire or resistors will be frequency dependent.  This is because the grid size sets an upper limit to the frequency up to which the grid may be usefully approximated as a continuous sheet. The reflection and transmission coefficients for a soldered resistive wire grid has been computed by Astrakhan [13]. Using their relations, we plot in Fig.~\ref{fig:graph_gridloss} the gain loss of the zero-spacing interferometer arising from the use of a coarse grid for the resistive sheet.  It is seen that provided the wavelength of operation exceeds eight grid units, the loss in response is less than $10\%$.  

\subsection{Performance dependence on the polarization of the incident EM wave}

\begin{figure}
\centering
\includegraphics[width = 3.4in]{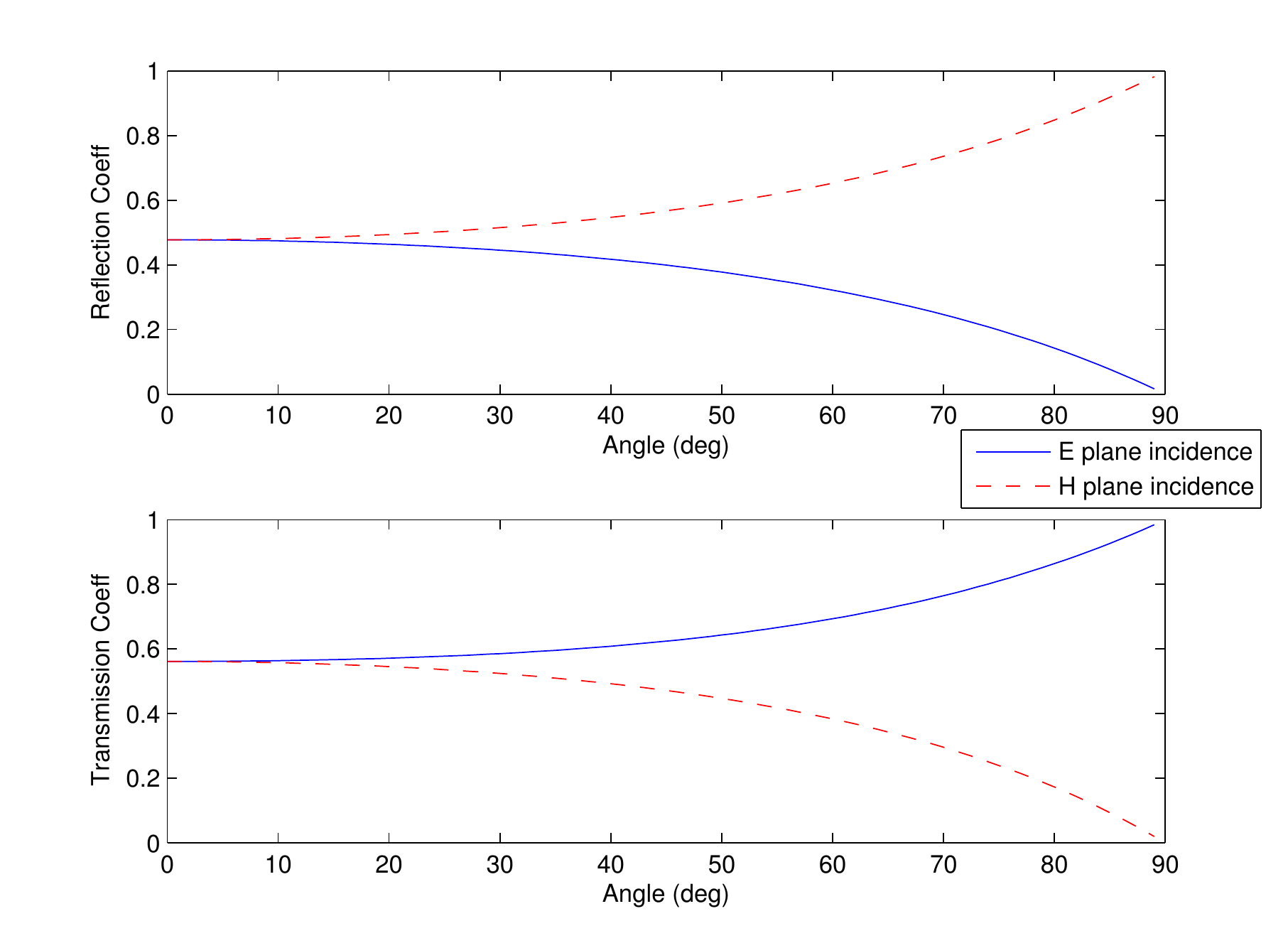}
\caption{Computed reflection and transmission coefficients for a wire grid, with element resistance equal to $\eta_o/2$, for varying angles of incidence. Separate curves show the coefficients for E-plane (polarization of the electric field parallel to the plane of incidence) and H-plane incidence (polarization perpendicular to the plane of incidence).}
\label{fig:graph_2}
\end{figure}

The values of the reflection and transmission coefficients of a wire grid were computed for different angles of incidence (Fig.~\ref{fig:graph_2}). The dependence of $\Gamma \tau$ product on the angle of incidence is same for both polarizations---the gain loss versus angle of incidence is independent of the plane of polarization---and for both polarizations the gain loss of the zero-spacing interferometer follows the $\Gamma \tau$ product given in Fig.~\ref{fig:angle_sensitivity}.

\subsection{Fabrication of a resistive sheet as a resistor grid}

As a proof of concept, we have constructed a resistive sheet beam splitter as a square grid of resistors.  Every segment of the soldered wire grid is made of conductive copper wire with a carbon resistor of value $180~\Omega$ at the center of each segment. This was the closest value to $\eta_o/2 = 188.5~\Omega$ that is readily available commercially. 

\begin{figure}
\centering
\includegraphics[width = 3.4in]{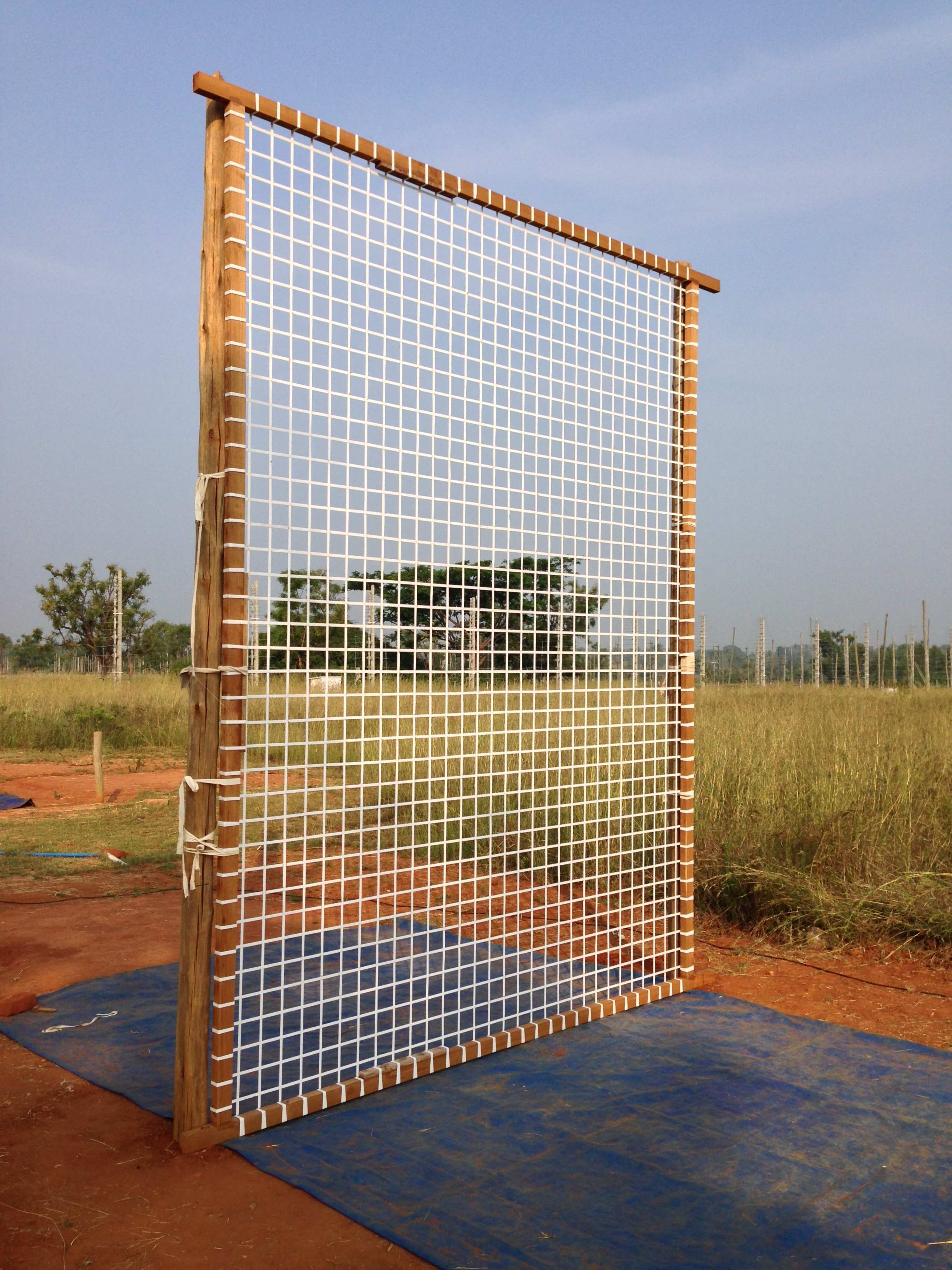}
\caption{Photograph of the resistor grid, which was constructed using a wooden frame, polypropylene strapping tape and 180~$\Omega$ resistors soldered at the centers of a 10-cm square conductive copper grid.}
\label{photo_1}
\end{figure}

\begin{figure}
\centering
\includegraphics[width = 3.4in]{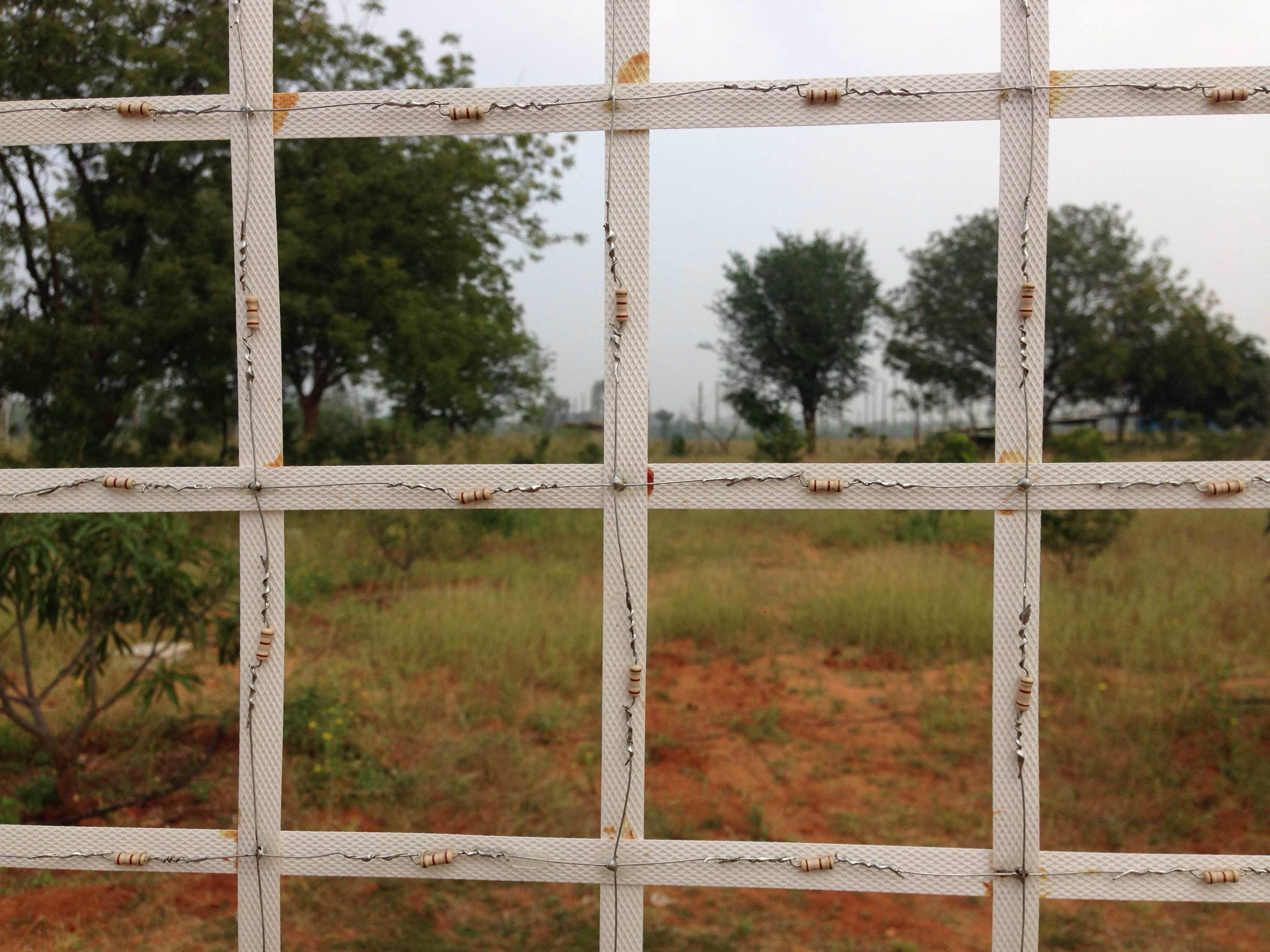}
\caption{A close up view of the resistor grid.}
\label{photo_2}
\end{figure}

The resistor grid was built across a 4~m $\times$ 3~m wooden frame that was constructed without using any conductive metal.  No nails or metal clips were used at joints.  The resistor network is supported by strapping tapes: a square grid of strapping tape was fastened on the wooden frame to serve as a base to support the resistor grid.  For support, a grid of tapes was used rather than a continuous sheet so that wind loading on the structure would be relatively smaller when deployed in the field. Strapping tape is commercially available as polypropylene (PP) and polyester (PET); we selected  PP strapping tape since the dielectric constant is 1.5--2.5 and close to unity compared to the 2.8--4.8 of PET.  The resistor grid network is formed by first making a square grid of copper wires over the PP strapping tapes, then soldering the wires at the junctions, then soldering a resistor parallel to the wires at the center of each wire segment and finally cutting away the shorting lengths of copper wire across each resistor to leave a resistor grid.  The resistors and soldered junctions are glued to the PP grid, which serves as a base that supports the resistor grid.  To enable the grid to have frequency independent characteristics up to about 350~MHz, the grid spacing is chosen to be 10~cm, which corresponds to $\lambda/8$ at 375~MHz.   Photographs of the prototype resistor-grid type EM beam splitter is shown in Fig.~\ref{photo_1} and a close-up of the grid is shown in Fig.~\ref{photo_2}.


\subsection{Measurements of transmission and reflection coefficients: comparison with modeling}

The beam splitter that we have constructed as a resistor grid was evaluated for its transmission and reflection properties at a number of discrete frequencies. We made measurements over the range 50 to 250~MHz in which the wavelength is greater than the grid cell size but also not much larger than the total dimensions of the resistor grid.  Transmission and reflection coefficients were measured using an Agilent FieldFox Network analyzer (N9915A) that was configured as a 2-port network analyzer and the coefficients were measured as $S_{11}$ and $S_{21}$ scattering parameters.  Linearly polarized monopole antenna elements were used, with lengths tuned to each of the discrete measurement frequencies.  
Measurements were compared with expectations based on modeling.

\begin{figure}
\centering
\includegraphics[width = 3.6in]{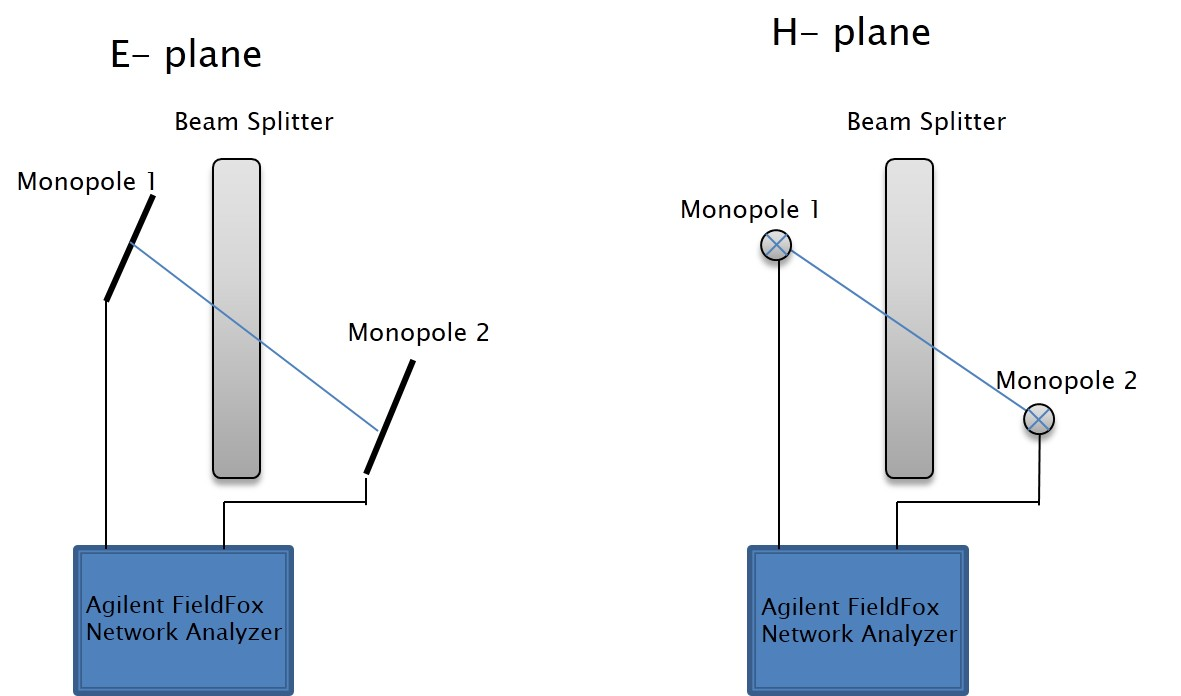}
\caption{An illustration of the setups used for the primary measurement of the transmission coefficient in the case of E-plane and H-plane incidence. The Network analyzer was configured in $S_{21}$ mode. In each case a secondary calibration measurement was done with the screen removed. For the measurement of reflection coefficients, a similar setup was used but with both the antennas placed on the same side of the resistor grid. A perfect reflector (Aluminum sheet) was used in place of the screen for additional calibration of the reflection characteristics.}
\label{schematic}
\end{figure}

The transmission measurements were made using linearly polarized antennas, one connected to the output terminal of the network analyzer on one side of the resistor grid and the second placed on the other side and connected via a coaxial cable to the input terminal of the analyzer.  The resistor grid was mounted vertically between the two antennas. A schematic of the set up in shown in Fig \ref{schematic}.  Each measurement was made with the antenna lengths adjusted to be resonant at the measurement frequency. The magnitude and phase of $S_{21}$  was recorded in a narrow bandwidth about the resonant frequency.  The $S_{21}$ measurement requires to be corrected for the return loss characteristics of the antennas used and also the space loss in order to derive the transmission coefficient of the resistor grid alone.  Therefore, additional calibration measurements were made. The measurement of $S_{21}$ was carried out for two cases, one primary measurement was made with the resistor grid between the transmitter and receiver antennas and a second calibration measurement with the resistor grid removed but with everything else in the setup and environment unchanged.  Calibration involves making two measurements, which we refer to as $S_{21}^{\rm on}$ and $S_{21}^{\rm off}$ respectively for the measurements made with and without the resistor grid between the antennas.  In terms of the corresponding magnitudes $V_{S_{21}^{\rm on}}$ and $V_{S_{21}^{\rm off}}$ and phases $\phi_{S_{21}^{\rm on}}$ and $\phi_{S_{21}^{\rm off}}$, the transmission coefficient is estimated as the ratio
\begin{equation}
\tau_{m} = \frac{ V_{S_{21}^{\rm on}}  e^{-j \phi_{S_{21}^{\rm on}}} }  { V_{S_{21}^{\rm off}}  e^{-j \phi_{S_{21}^{\rm off}}} }.
\label{tau_measurement}
\end{equation}
The ratio of the magnitudes of the pair of $S_{21}$ measurements yields the magnitude of  $\tau_m$ and the difference in the measured phases yields the phase of the transmission coefficient. Such a referenced measurement takes into account the gains of the antennas used, corrects for space propagation, and also cancels response arising from unwanted propagation paths via ground reflections and reflections off other structures in the surroundings.

For the measurement of $\tau_m$ at normal incidence, both the linearly polarized antennas were mounted with the electric field polarization direction perpendicular to ground.  The antenna phase centers were kept at a common height of 2~m above ground, which is half the total height of the $4 \times 3$~m$^2$ resistor grid.  Measurements of $\tau_m$ were also made for oblique incidence of $30^{\circ}$ by raising the transmitter antenna and lowering the receiving antenna so that the straight ray propagation path intersected the resistor grid close to its center and with this incidence angle of $30^{\circ}$.  For oblique incidence, measurements of $\tau_m$ were made separately for $H$-plane incidence, in which the transmitter and receiver antennas were mounted parallel to ground, and for $E$-plane incidence, in which the antennas were mounted in the plane of incidence and inclined at $30^{\circ}$ to the vertical. At each frequency, the distance between the antennas and resistor grid was adjusted so that the Fresnel number $d^2/(L \times \lambda)$ was in the range 0.3--0.5, where $d$ is the dimension of the resonant antenna, $\lambda$ is the wavelength of  the measurement and $L$ is the distance between the antennas and the resistor grid.

\begin{figure}
\centering
\includegraphics[width = 3.6in]{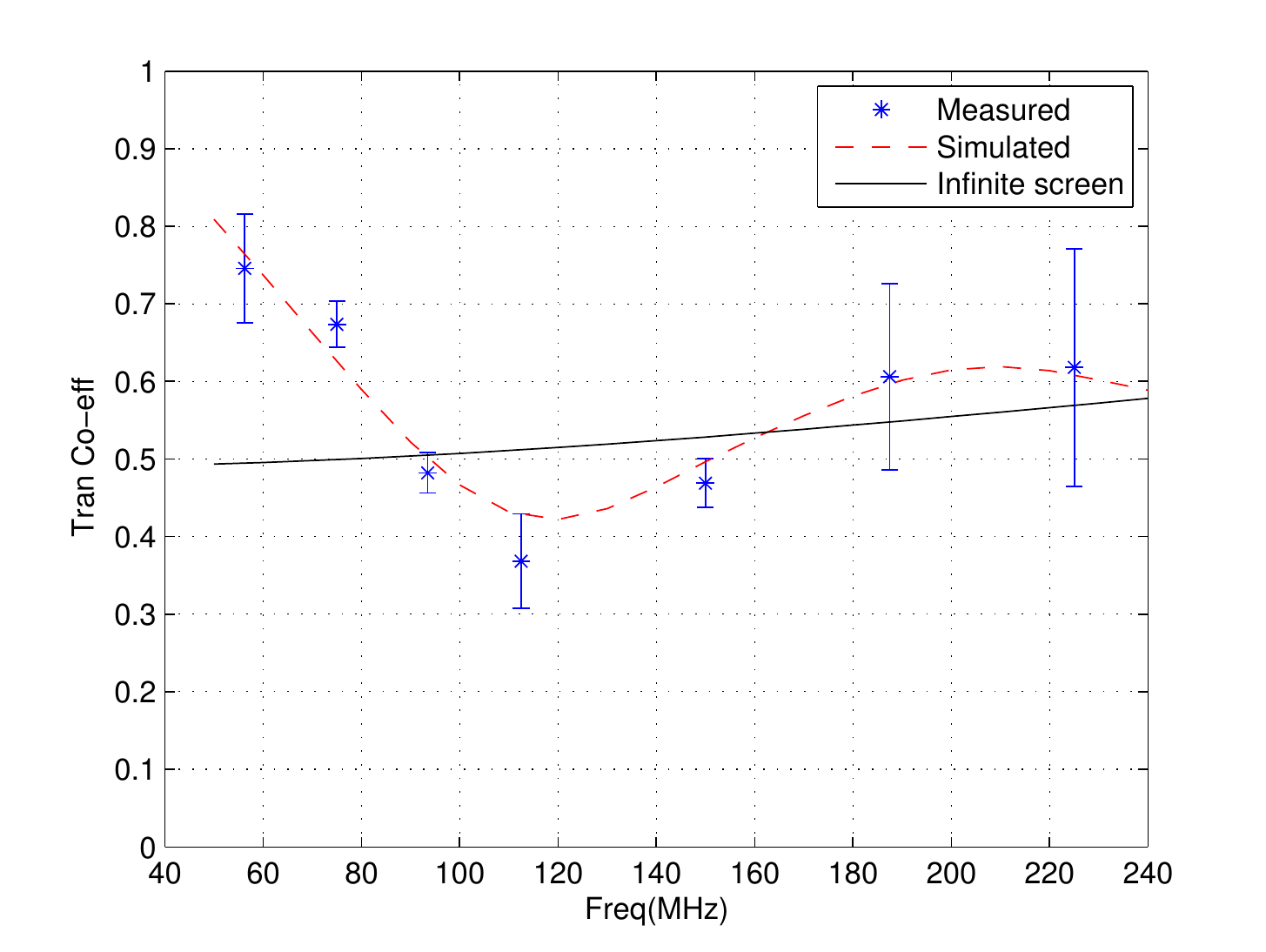}
\caption{$\tau$ versus frequency for the resistor-grid type space beam splitter for normal incidence.  The measured $\tau_m$ are shown using symbols.  The expectations for $\tau$ for the case of an infinite resistor grid is shown as a continuous black line. Also shown  for comparison using a  red dashed line is the expectation based on physical optics modeling of the transmission, which was done specifically for a finite size resistor grid of $4 \times 3$~m$^2$ that is the same as the one on which the measurements were made.  }
\label{graph_3}
\end{figure}

\begin{figure} 
\centering
\includegraphics[width = 3.6in]{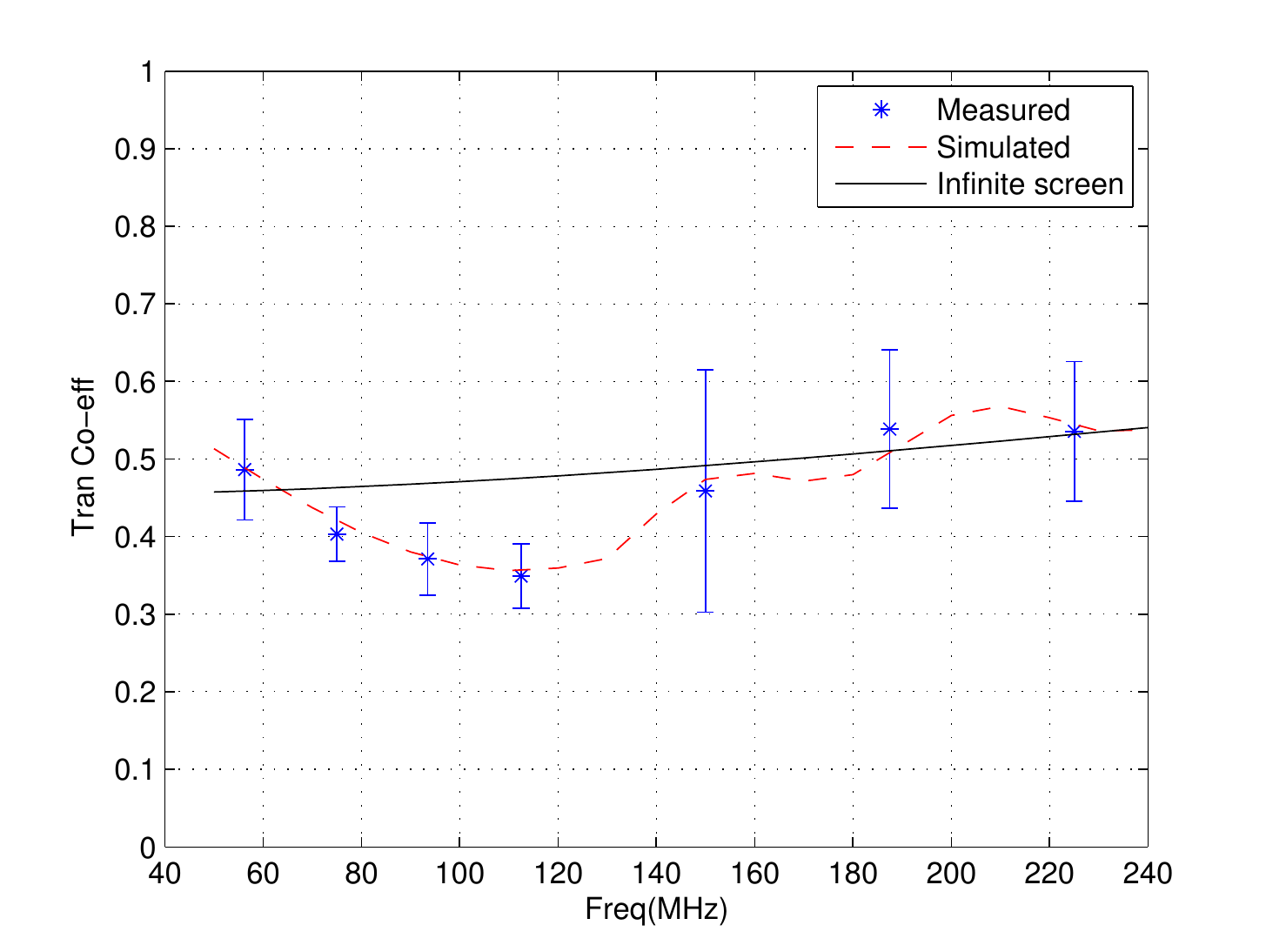}
\caption{$\tau$ versus frequency for H plane incidence at $30^{\circ}$. As in the previous figure, comparisons are provided for corresponding expectations for finite size and infinite resistor grids.}
\label{graph_4}
\end{figure}

\begin{figure} 
\centering
\includegraphics[width = 3.6in]{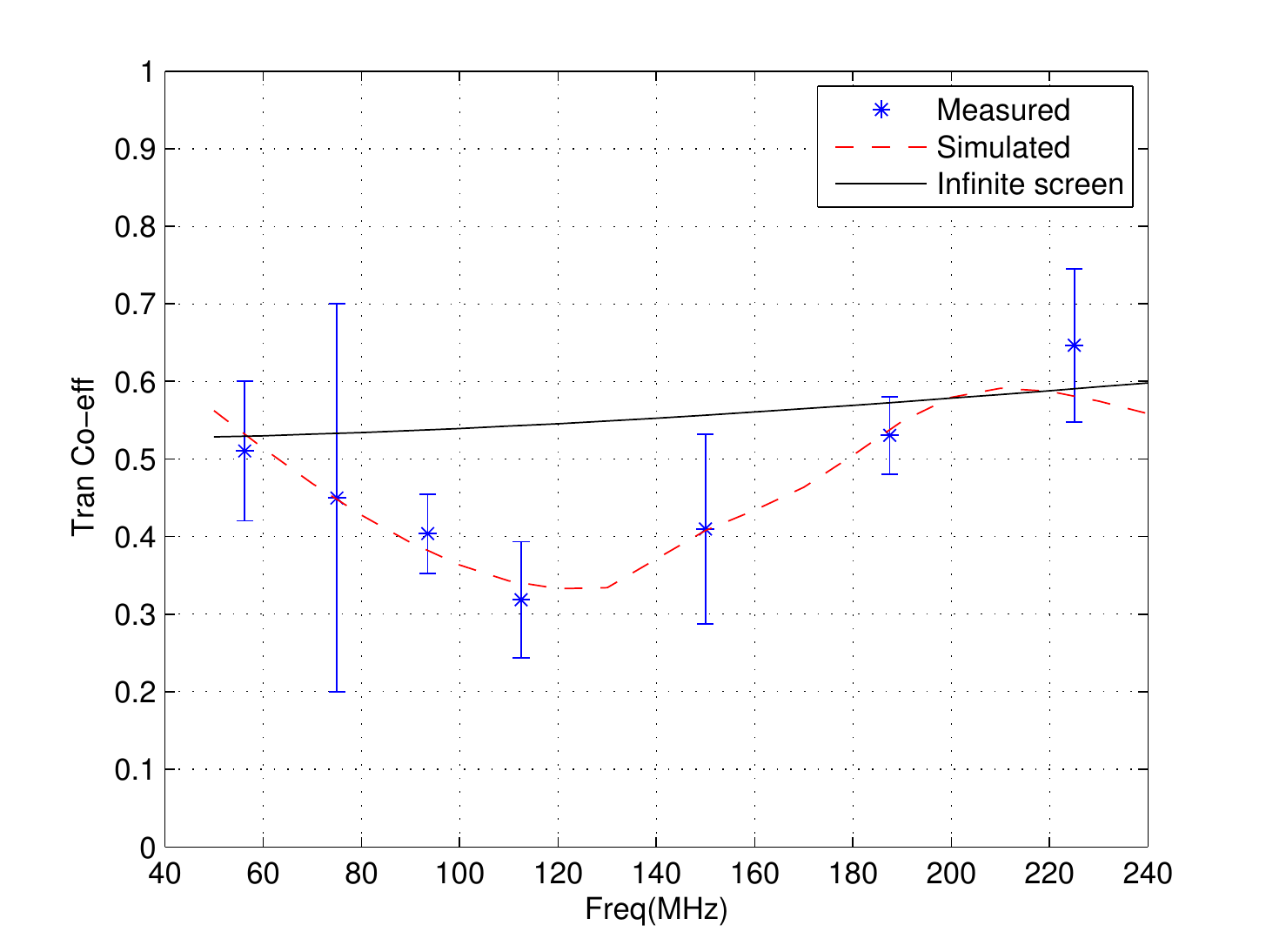}
\caption{$\tau$ versus frequency for E plane incidence at $30^{\circ}$. As in the previous two figures, comparisons are provided for corresponding expectations for finite size and infinite resistor grids.}
\label{graph_5}
\end{figure}

Fig.~\ref{graph_3} shows the measured transmission coefficient $\tau_m$ versus frequency for normal incidence.  Plots of $\tau$ versus frequency for oblique incidence at  $30^{\circ}$ are shown in Figs.~\ref{graph_4} and \ref{graph_5} separately for the two cases where the incident electric field is in the $H$ and $E$ plane respectively.  Also shown, for comparison, are the expectations for $\tau$ based on Astrakhan [13].

Measurements of reflection coefficients were also carried out for three cases: normal incidence, H plane incidence at an angle of $30^{\circ}$ and E plane incidence at an angle of $30^{\circ}$. For normal incidence, a single antenna was used as a trans-receiver.  The trans-receiver antenna was placed on one side of the resistor grid and connected to the output terminal of the network analyzer. In this particular case, the $S_{11}$ scattering parameter served as a measurement of the reflected power.  The linearly polarized antenna was mounted perpendicular to the ground at a fixed height of 2.0~m and at each measurement frequency the distance between the antenna and resistor grid was adjusted to be four times the Fresnel number. Oblique incidence measurements were made separately using a pair of antennas and as the $S_{21}$ scattering matrix element.  For these reflection coefficient measurements both antennas were placed  on the same side of the resistor grid.  As in the transmission case, for H-plane measurements the linearly polarized antennas were mounted horizontally and their relative heights above ground were adjusted so that the nominal ray path between the antennas reflected off the center of the screen with an incidence angle of  $30^{\circ}$. The pair of antennas were placed in the plane of incidence for E-plane measurements, with one tilted to $30^{\circ}$ and the other to $270^{\circ}$ with respect to the vertical.

Calibration of the reflection measurements to derive the reflection coefficients of the resistor grid required additional calibration measurements.  These secondary measurements of the scattering matrix element were done separately without the grid and, additionally, with a reflector plate (aluminum sheet) replacing the resistor grid. The amplitude and phase recorded in this second calibration measurement are denoted by  $V_{S_{ ij}^{\rm al}}$ and  $\phi_{S_{ ij}^{\rm al}}$, where $ij$ is either 11 or 21 depending on whether $S_{11}$ or $S_{21}$ was measured. The complex $S_{11}^{\rm off}$ and $S_{21}^{\rm off}$ calibration measurements that were made without the resistor grid in place were subtracted from the corresponding measurements made with the resistor grid and with aluminum reflector in place.  This subtraction was aimed at cancellation of space propagation terms and invariant extraneous reflections from the environment. The reflection coefficient of the resistor grid is estimated as follows:
\begin{equation}
\Gamma_{m} = \frac{ V_{S_{ij}^{\rm on}}  e^{-j \phi_{S_{ij}^{\rm on}}} - V_{S_{ij}^{\rm off}}  e^{-j \phi_{S_{ij}^{\rm off}}}}  
{ V_{S_{ij}^{\rm al}}  e^{-j \phi_{S_{ij}^{\rm al}}} - V_{S_{ij}^{\rm off}}  e^{-j \phi_{S_{ij}^{\rm off}}} }.
\label{gamma_measurement}
\end{equation}

\begin{figure} 
\centering
\includegraphics[width = 3.6in]{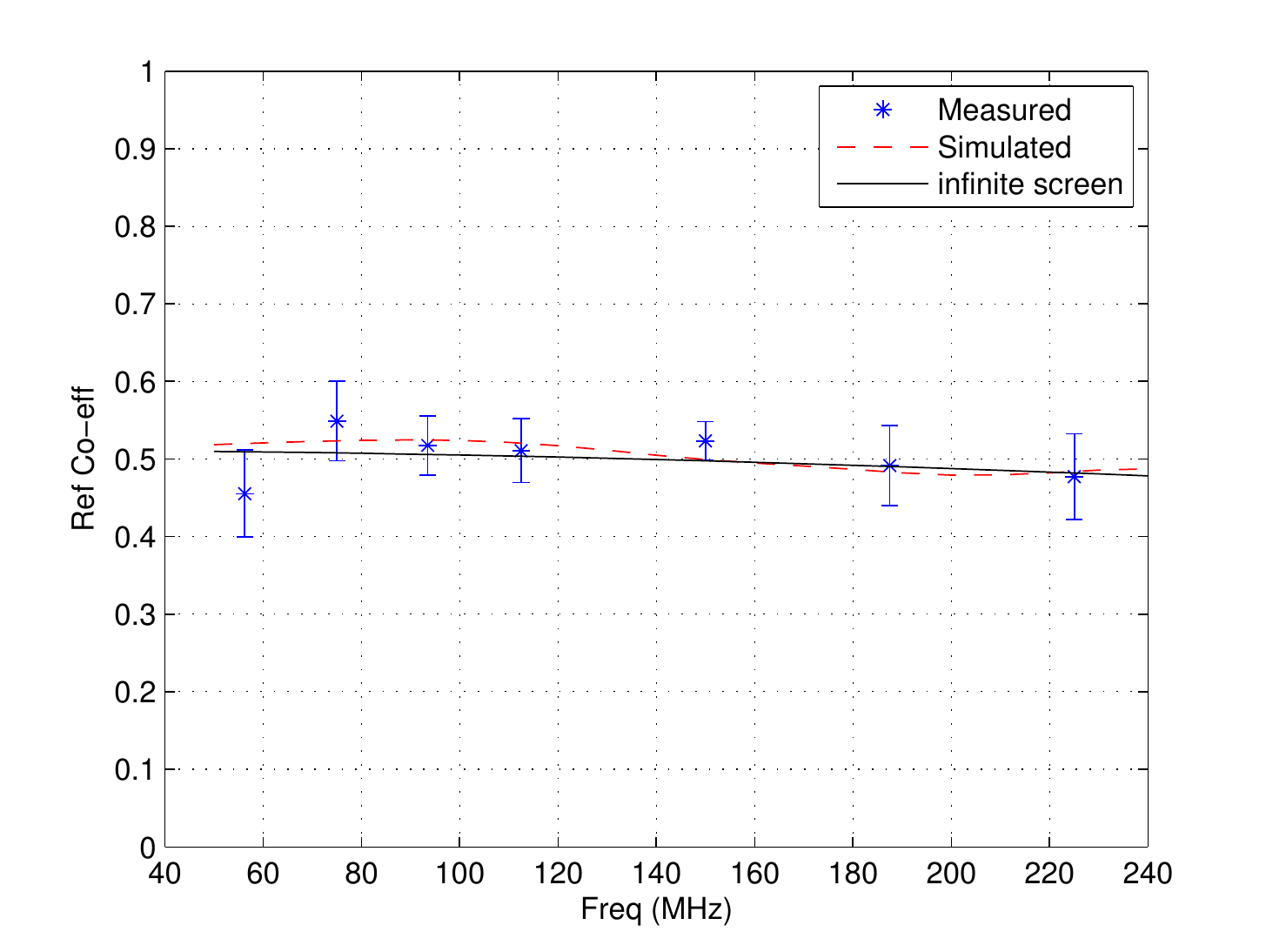}
\caption{$\Gamma$ versus frequency for the resistor grid type space beam splitter for normal incidence. 
As in the previous figures, measurements are compared with corresponding expectations for finite size and infinite resistor grids.}
\label{graph_6}
\end{figure}

\begin{figure}
\centering
\includegraphics[width = 3.6in]{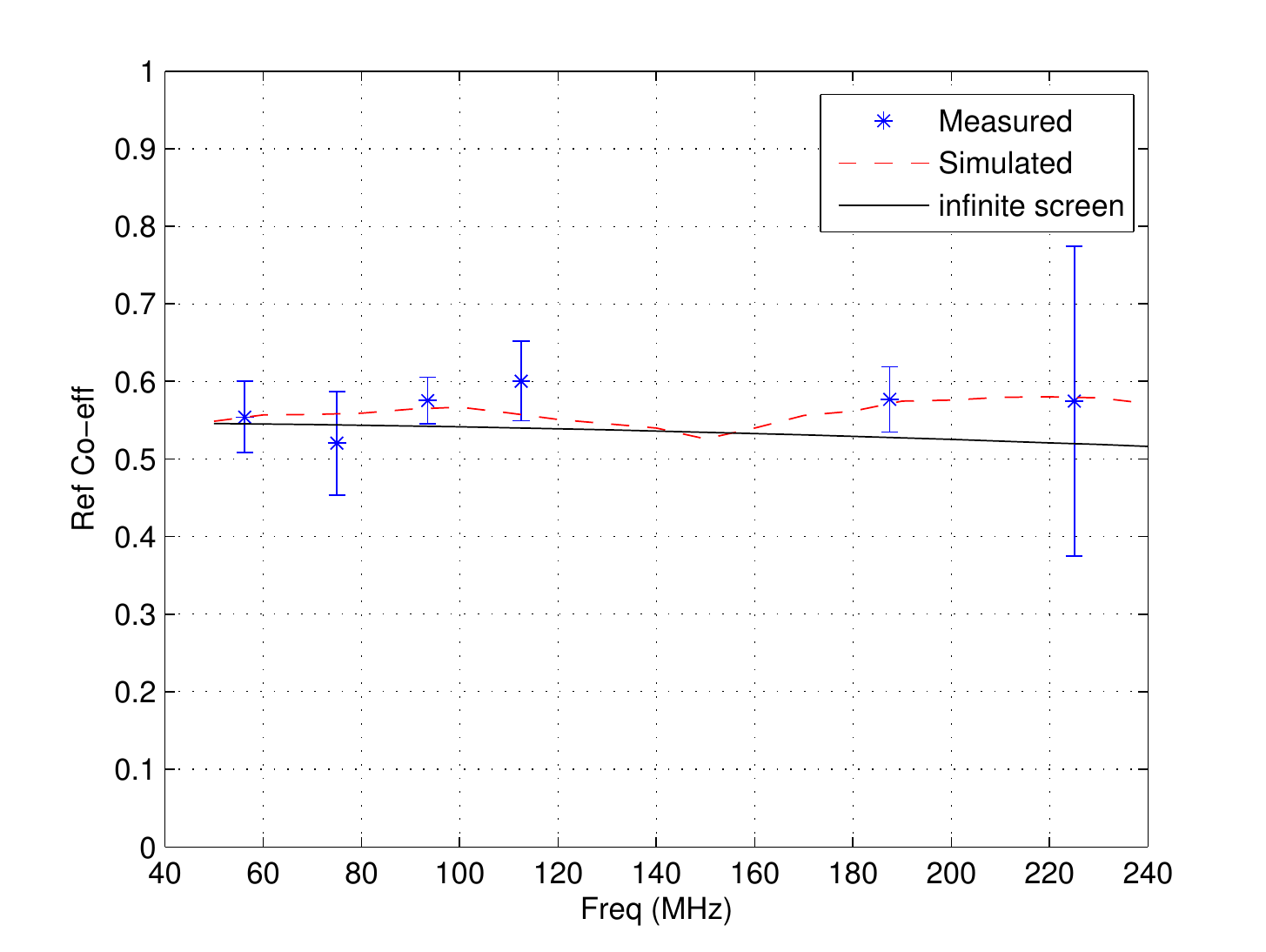}
\caption{$\Gamma$ versus frequency for H plane incidence at $30^{\circ}$. As in the previous figures, measurements are compared with corresponding expectations for finite size and infinite resistor grids.}
\label{graph_7}
\end{figure}

\begin{figure} 
\centering
\includegraphics[width = 3.6in]{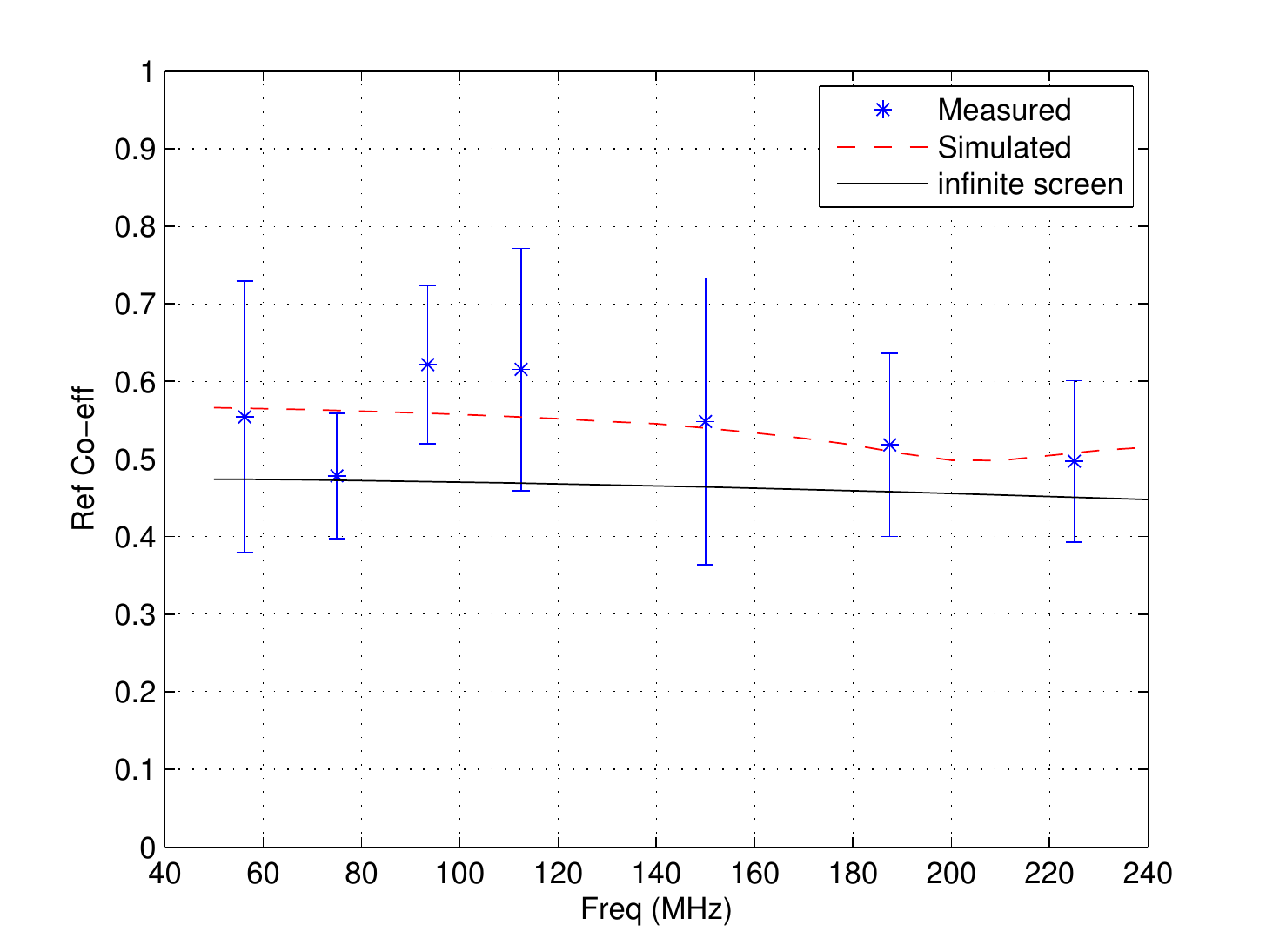}
\caption{$\Gamma$ versus frequency for E plane incidence at $30^{\circ}$. As in the previous figures, measurements are compared with corresponding expectations for finite size and infinite resistor grids. }
\label{graph_8}
\end{figure}

Fig.~\ref{graph_6} shows a plot of reflection coefficient $\Gamma_{m}$ versus frequency for normal incidence.  Plots of $\Gamma_{m}$ versus frequency for oblique incidence at $30^{\circ}$ are shown in Figs.~\ref{graph_7} and \ref{graph_8} for cases where the incident electric fields are, respectively, in the $H$ and $E$ planes.  

As expected, the modeling of the reflection and transmission that takes into account the finite size of the resistor grid provides a better match to the measurements.  These refined computations of the expected coefficients are based on physical optics modeling of the diffraction along the edges.  The numerical computation sampled the electric field over a finely spaced square array of points on the plane of the resistor grid. The sampling extended over an area up to ten times larger in extent along the upper and two adjacent sides of the resistor grid. The bottom fourth side was omitted since ground blocks paths in that dimension. The first step of the computation was to propagate all rays from the phase center of the transmitter antenna to each and every point on the chosen grid plane assuming inverse square law propagation of power in free space. The computed electric field at each sampling point on the plane was computationally reflected or transmitted knowing the angle of arrival of each ray.  Reflection and transmission coefficients of the resistor grid were adopted over the  $4 \times 3$~m$^2$ area of the resistor grid.   Beyond that, transmission coefficient of unity and reflection coefficient of zero was assumed. In the case of E-Plane incidence in which the electric field is oriented in the plane of incidence, a correction factor of (cos$\theta \times {\rm cos}\theta$) was applied to account for the tilt between the measuring antennas and the plane of the resistor grid. The array of emergent fields were propagated with space loss to the receiving antenna and the components along the polarization direction of the receiving antenna were added vectorially to obtain the net response.

In the case of transmission, the numerical computation of the net received field is repeated for the case where there is no resistor grid between the antennas.  The pair of computations made for the cases with and without the resistor grid are used as in Equation~\ref{tau_measurement} to derive the expectation for $\tau$. The computation of the net received field for reflection is repeated for the case where the resistor grid is replaced with a perfect reflector with $\Gamma = -1$ and the pair of computations used as in Equation~\ref{gamma_measurement} to derive the expectation for reflection coefficient.  

The results of this physical optics based expectation for the measurements of transmission and reflection coefficients are also shown in Figs.~\ref{graph_3}--\ref{graph_8} along with expectations corresponding to an infinite size for the resistor wire grid. If an infinite screen with ($\eta_o/2)~\Omega$~square$^{-1}$ sheet resistance were used for power splitting, then the reflection and transmission coefficients would be close to 0.5. The physical optics modeling demonstrates good agreement between measurements and modeling for the finite size resistor grid based space beam splitter: the discordance between modeling and measurements has a standard deviation of about 5\%.

\begin{figure}[h!] 
\centering
\includegraphics[width = 3.6in]{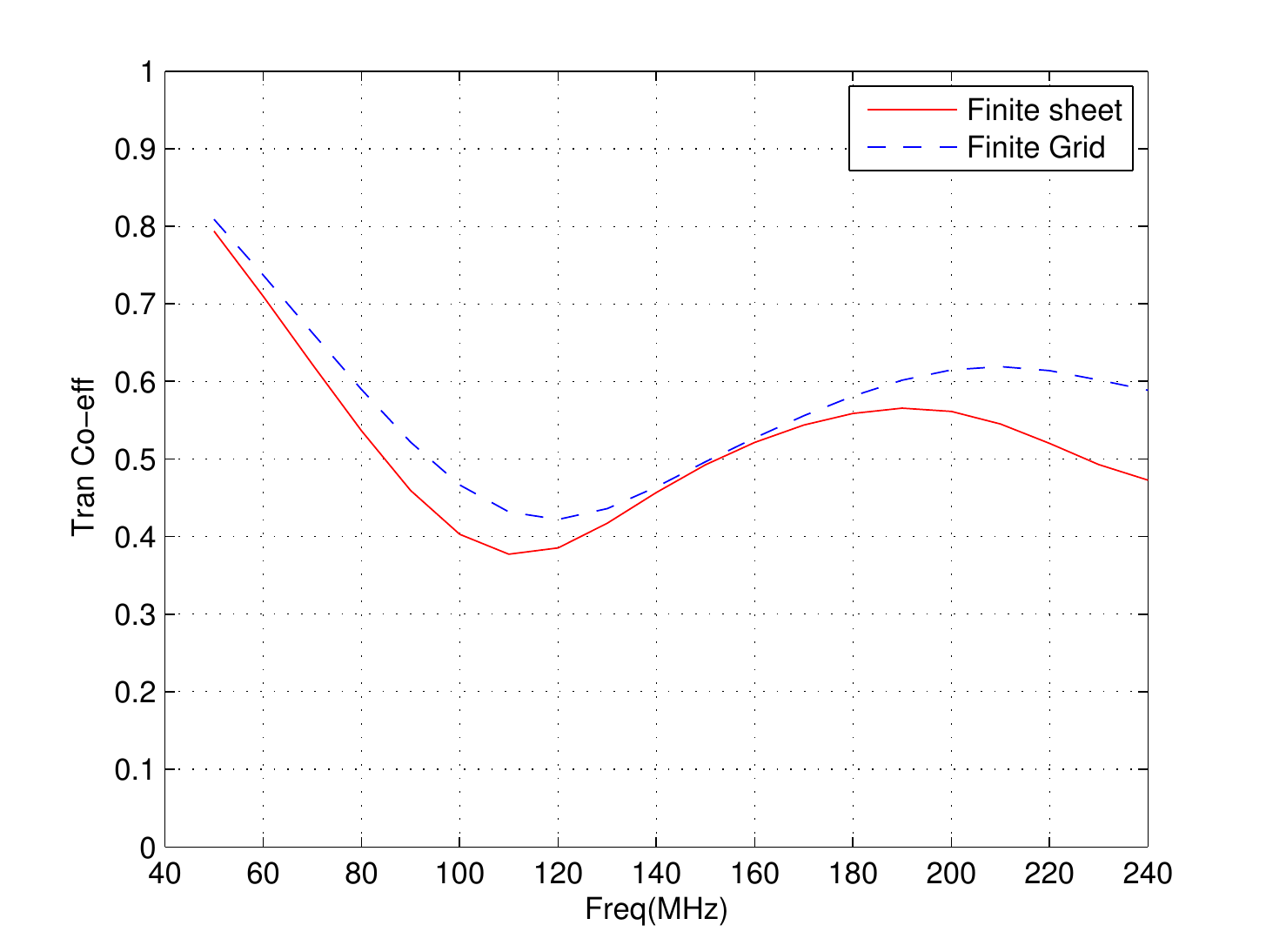}
\caption{ $\tau$ versus frequency for normal incidence on a $4 \times 3$~m$^2$ resistive sheet type space beam splitter (red line).  In this figure and in all following figures, also shown for comparison is the expectations for the case of a resistor grid of $4 \times 3$~m$^2$ (blue dashed line).}
\label{graph_9}
\end{figure}
\clearpage
\begin{figure}[h!]
\centering
\includegraphics[width = 3.6in]{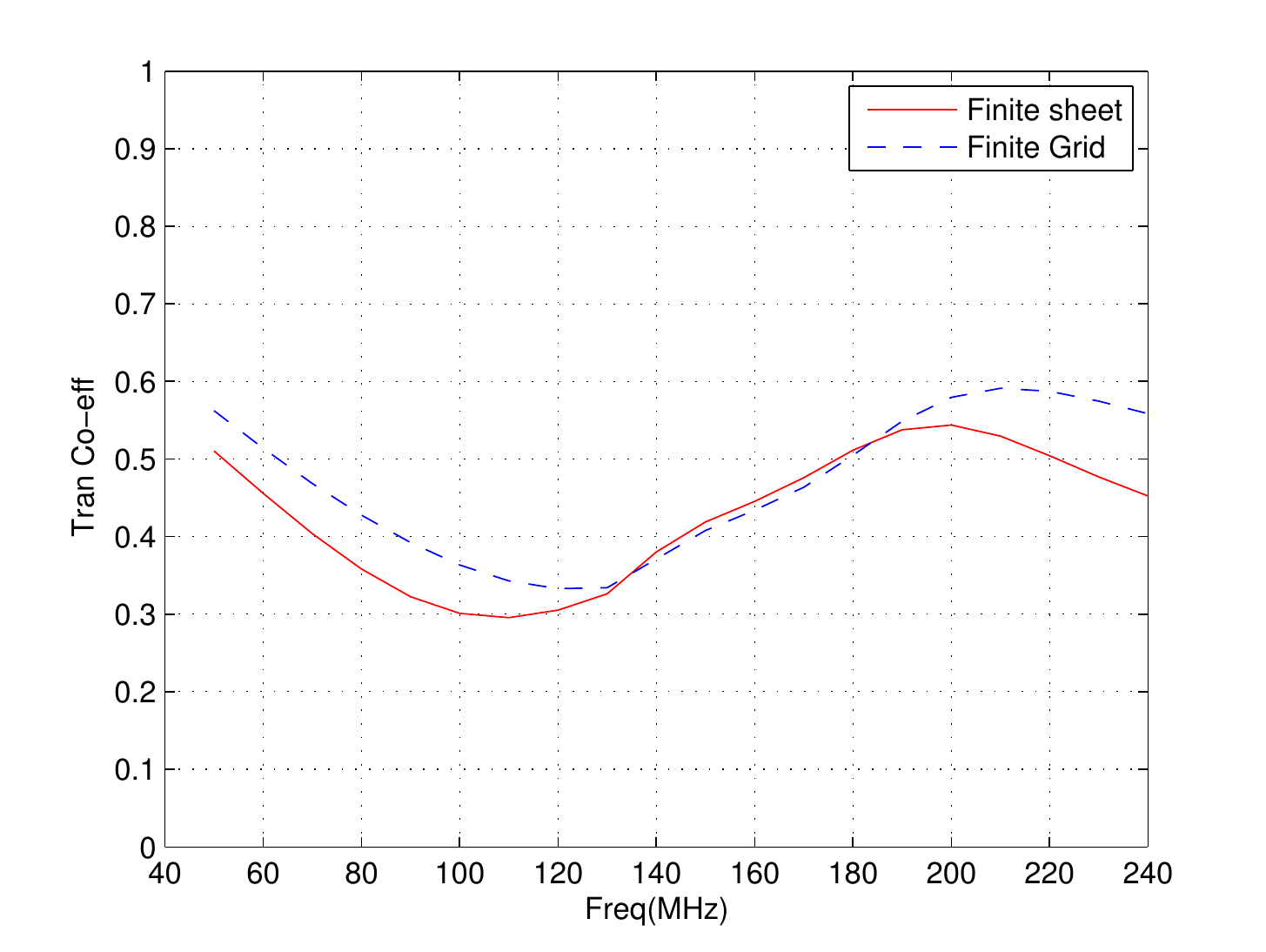}
\caption{$\tau$ versus frequency for E plane incidence at $30^{\circ}$. }
\label{graph_10}
\end{figure}

\begin{figure}[h!] 
\centering
\includegraphics[width = 3.6in]{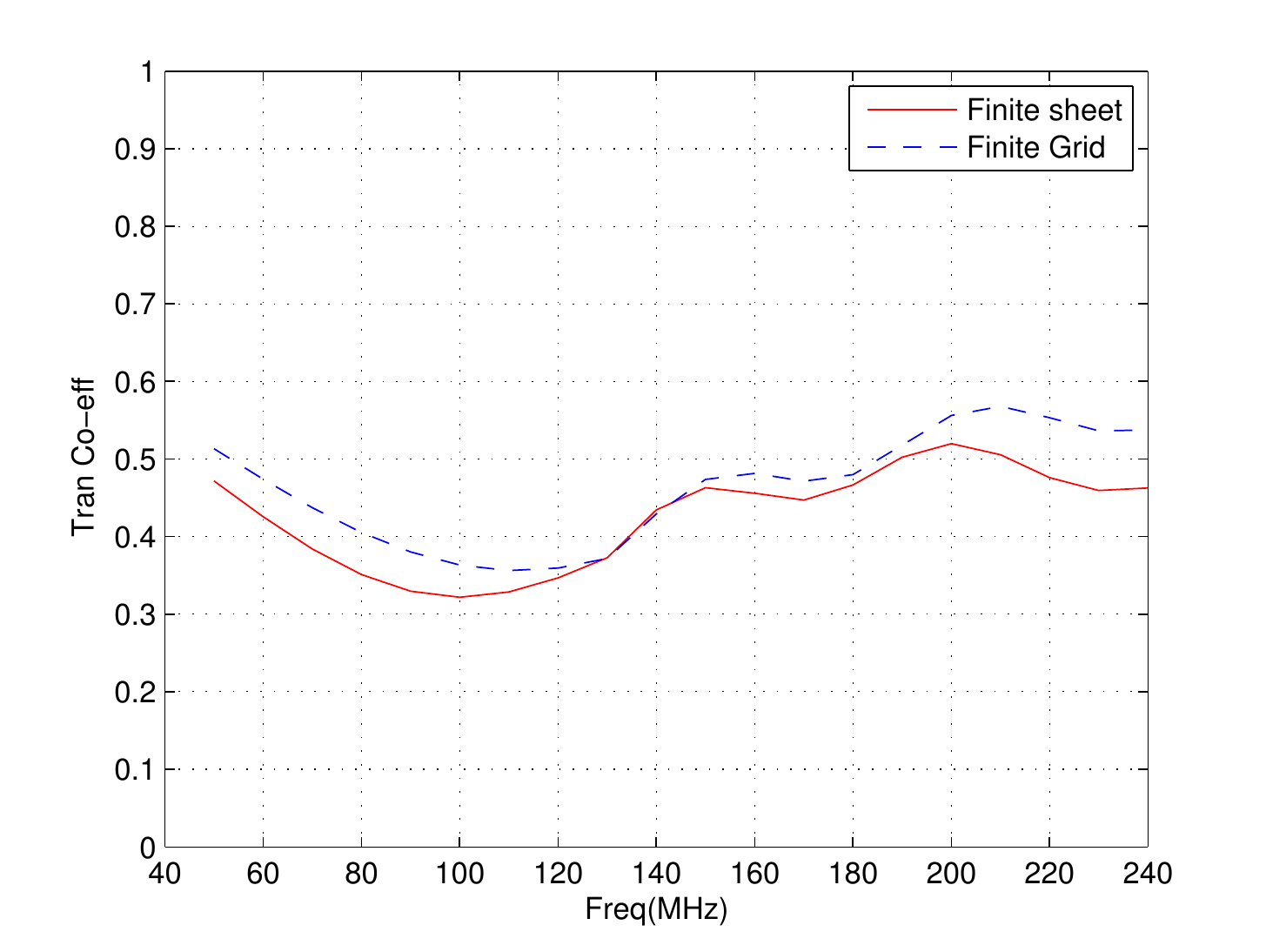}
\caption{$\tau$ versus frequency for H plane incidence at $30^{\circ}$.  }
\label{graph_11}
\end{figure}

A resistor wire grid is a good approximation to a resistive sheet and may be made arbitrarily close in performance to a sheet by reducing the grid size.  From the point of view of minimizing wind loading on a vertically mounted beam splitter of the dimensions required for meter wavelength operation, a resistor wire grid may be a practically preferred alternative to a resistive sheet.  However, a resistive sheet would have wider bandwidth of operation compared to a grid and the sheet would be frequency independent over the entire domain wherein the sheet resistance remains close to the nominal value.  However, in both cases, the finite size of realizations of space beam splitters would manifest edge diffraction effects in the response, which would be frequency dependent.

\begin{figure}[h!]
\centering
\includegraphics[width = 3.6in]{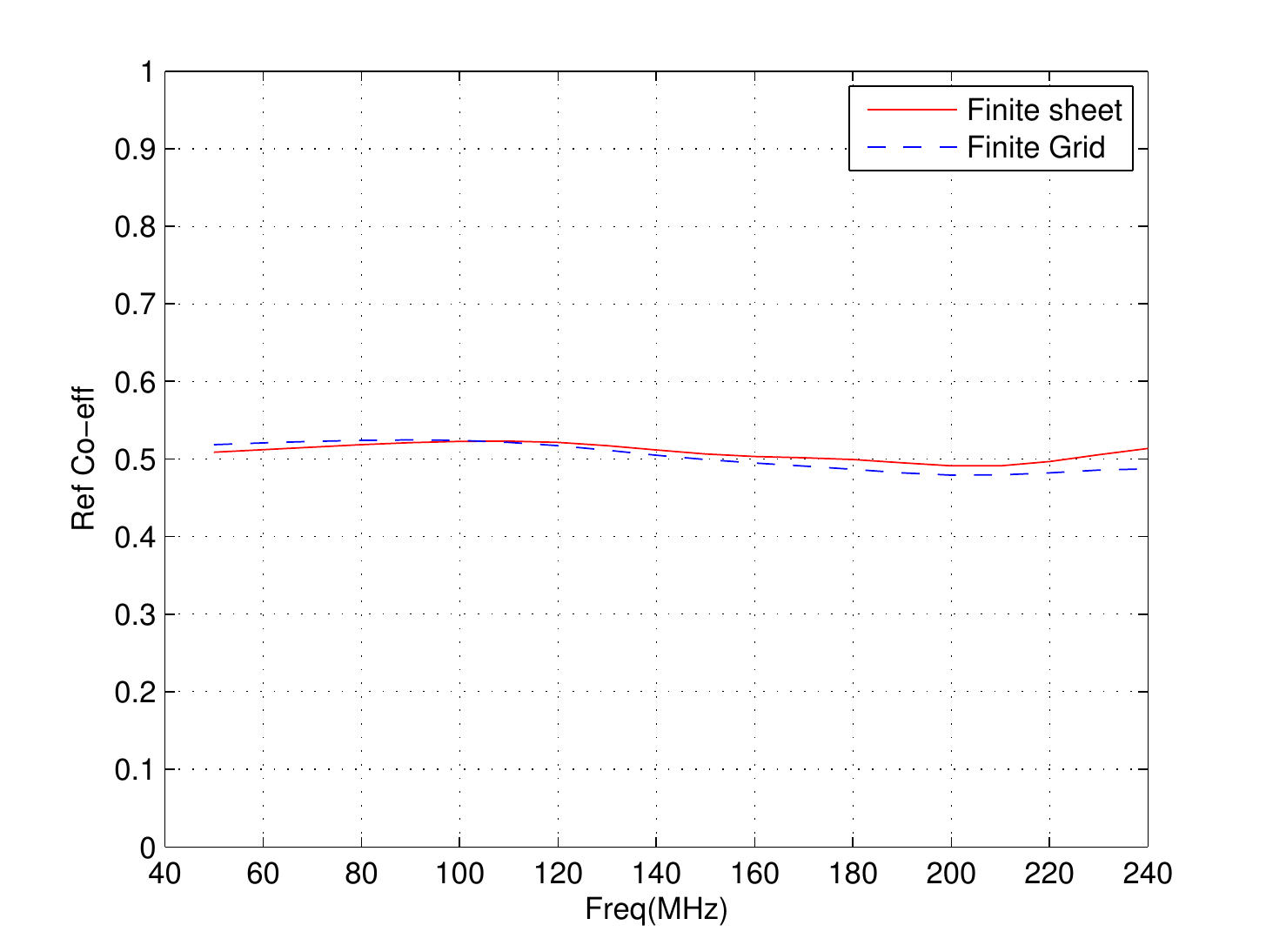}
\caption{ $\Gamma$ versus frequency for normal incidence. }
\label{graph_12}
\end{figure}

\begin{figure}[h!] 
\centering
\includegraphics[width = 3.6in]{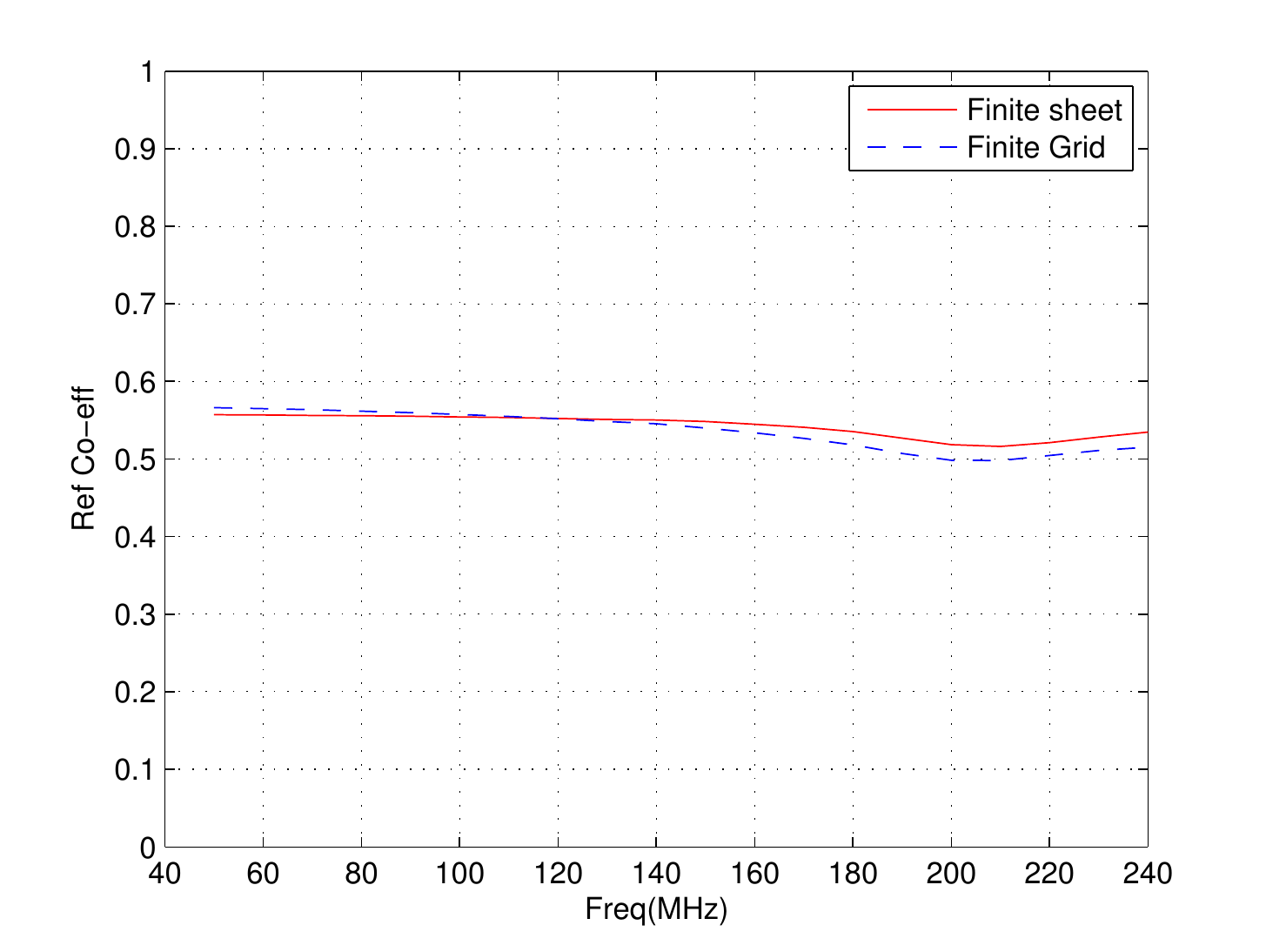}
\caption{$\Gamma$ versus frequency for E plane incidence at $30^{\circ}$. }
\label{graph_13}

\end{figure}

\begin{figure}[h!]
\centering
\includegraphics[width = 3.6in]{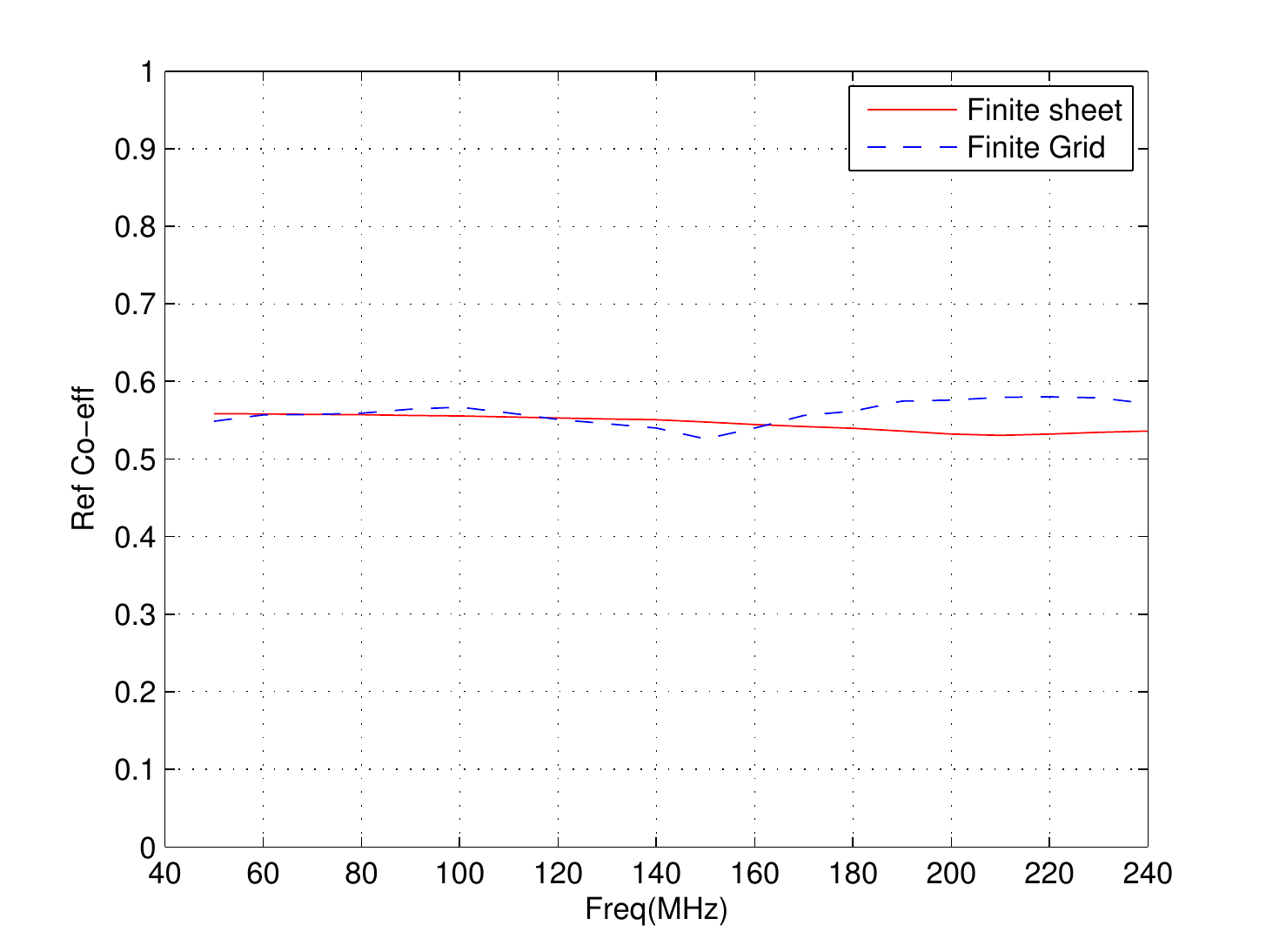}
\caption{$\Gamma$ versus frequency for H plane incidence at $30^{\circ}$. }
\label{graph_14}
\end{figure}

As a comparison and reference, we have also computed the transmission and reflection coefficients for a $4 \times 3$~m$^2$ beam splitter that is assumed to be a resistive sheet of ($\eta_o/2)~\Omega$~square$^{-1}$ sheet resistance.   These are shown in Figs.~\ref{graph_9}--\ref{graph_14} in which the response of such a finite size resistive sheet is compared with a resistor grid of the same total size; it may be noted that the resistor grid is assumed to have resistors of value $180~\Omega$ that is somewhat different from $(\eta_o/2)~\Omega$. The figures show that the electromagnetic performances of the wire grid and continuous sheet beam splitters are very similar in the frequency range explored here, in which the grid size is a small fraction of the wavelength.

\section{Summary}

Events in the early history of the Universe are believed to have left imprints in the spectrum of the cosmic radio background.  Their detection requires precision receivers.  A major problem in their detection is confusion from internal noise of spectral radiometers, which includes self-generated noise in the low-noise amplifiers as well as noise added by ohmic losses in the antenna and passive interconnects.  Radiometers operating as interferometers avoid this problem since ohmic losses and amplifier noise are uncorrelated between elements in different arms. However, detection of uniform sky brightness with an interferometer requires a space beam splitter operating at radio wavelengths.

Towards building such a zero-spacing interferometer for absolute measurements of spectral signatures in the radio sky, we have developed an electromagnetic space beam splitter as a semi-transparent resistive sheet.  A two-element interferometer with antennas on either side of a vertical resistive sheet forms a zero-spacing interferometer.  The antenna beam solid angle is expected to be directed at the vertical resistive sheet so as to dominantly respond to sky brightness that is either reflected or transmitted from the sheet on the two sides.   With the use of the beam splitter, half the incident sky power is absorbed in the resistive sheet, a quarter is transmitted and a quarter reflected.  Resistive sheets are frequency independent and hence their reflection ($\Gamma$) and transmission ($\tau$) properties are independent of frequency, thus providing smooth transfer function for the sky spectrum without altering the cosmological signatures embedded therein.  

The electromagnetic beam splitter has been realized as a square grid of resistive wire or a square grid of resistors.  Such a realization would, admittedly, have frequency independent characteristics only at wavelengths substantially longer than the grid size and considerably smaller than the physical size of the sheet.  Reflection and transmission coefficients were measured, both for normal as well as oblique incidences, as well as for cases where the incident electric field is in the plane of incidence and perpendicular to this plane.  The measurements were done as $S_{11}$ and $S_{21}$ scattering matrix parameters using a network analyzer and a pair of linearly polarized antennas resonant at the measurement frequency.  The measurements were calibrated using corresponding measurements made without the resistor grid and separately by placing a conductive aluminum sheet in its place.  The measurements of reflection and transmission coefficients agree with expectations.

The development continues with designing antennas appropriate for the zero-spacing interferometer configuration, which would then  lead to a system design.

\section*{Acknowledgment}

The work has been made possible by outstanding support from staff of the Gauribidanur Radio Observatory and the Electronics Laboratory and Mechanical Engineering services at the Raman Research Institute.

\begin{IEEEbiography}
[{\includegraphics[width=1in,height=1.25in,clip,keepaspectratio]{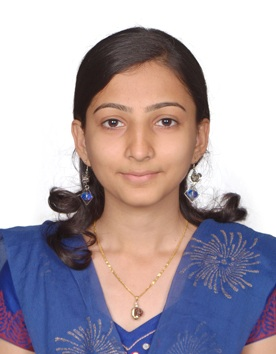}}]
{Nivedita Mahesh}
received the B.E degree in Instrumentation and Control Engineering from PSG College of Technology, Coimbatore, India, in 2012.
During 2012-14 she was with the Raman Research Institute, Bangalore,~India and is now a Masters student in the Antenna Research, Analysis, and Measurement Laboratory at UCLA.
\end{IEEEbiography}

\begin{IEEEbiography}
[{\includegraphics[width=1in,height=1.25in,clip,keepaspectratio]{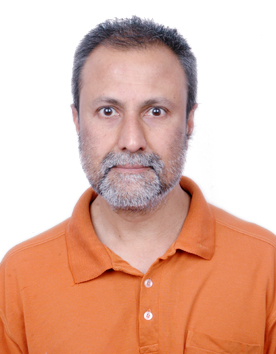}}]
{Ravi Subrahmanyan}
received the B.Tech. degree in Electrical Engineering from
the Indian Institute of Technology, Madras, India, in 1983
and the Ph.D. degree in astronomy from the Physics department of the
Indian Institute of Science, Bangalore, India, in 1990.
Since 2006, he has been at the Raman Research Institute, Bangalore, India. 
\end{IEEEbiography}

\begin{IEEEbiography}
[{\includegraphics[width=1.25in,height=1.75in,clip,keepaspectratio]{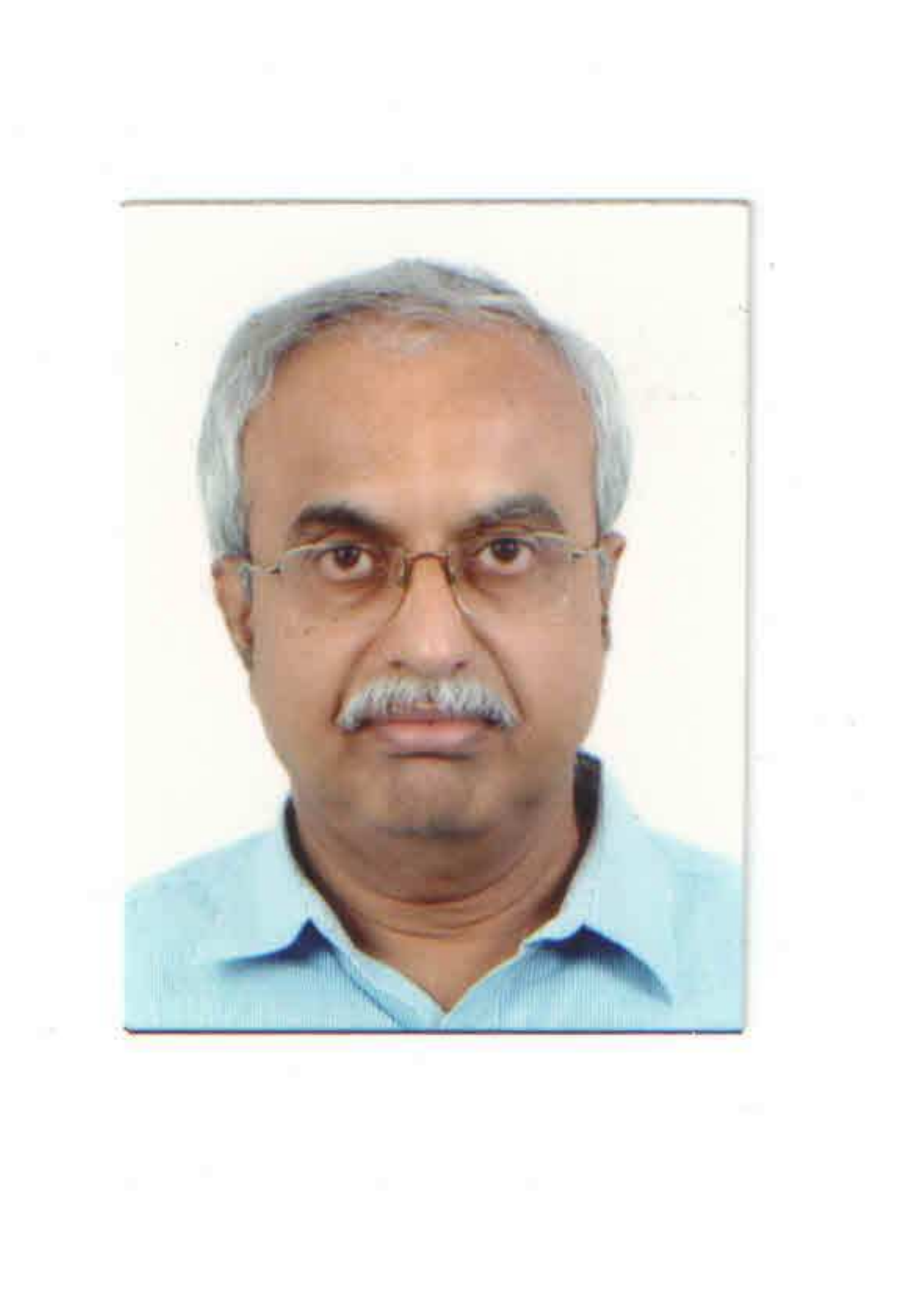}}]
{N. Udaya Shankar}
 received the M.Sc. degree in Physics from the Bangalore University in 1973 and Ph.D. degree in astronomy in 1986. Since 1978 he has been at the Raman Research Institute, Bangalore, India, working on the instrumentation for aperture arrays, wide-field imaging and sky surveys.
\end{IEEEbiography}

\begin{IEEEbiography}
[{\includegraphics[width=1in,height=1.25in,clip,keepaspectratio]{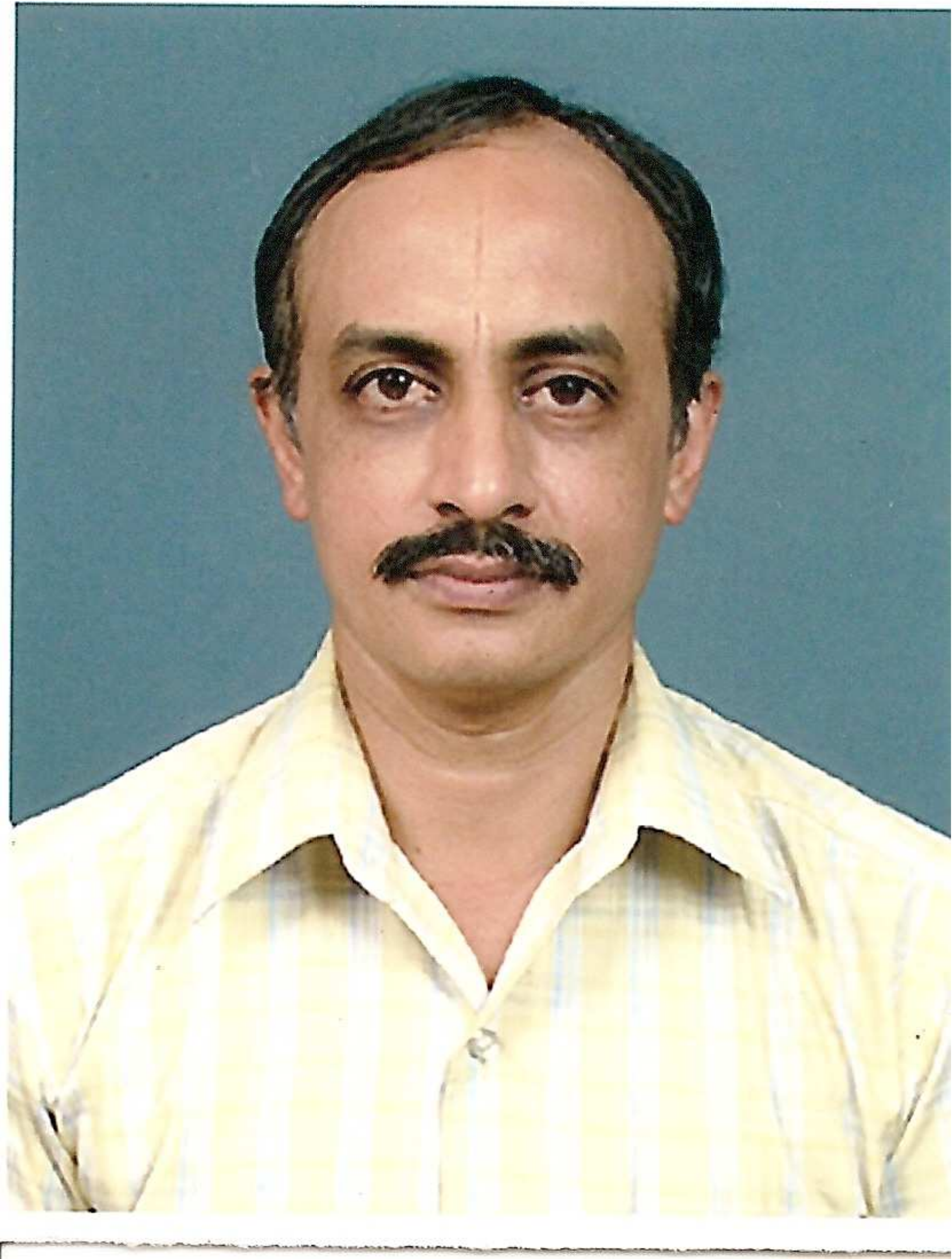}}]
{Agaram Raghunathan}
received the B.E. degree in Instrumentation Engineering from Bangalore Institute of Technology,
in 1990 and M.Sc. Engg. (by research) from the Engineering Department of the Bangalore
University, India, in 2000. Since 1990, he has been at the Raman Research 
Institute, Bangalore, India.
\end{IEEEbiography}

\vfill







\end{document}